\documentclass[12pt,fleqn]{article}
\usepackage{verbatim,amssymb,amsmath,graphics,amsthm}
\numberwithin{equation}{section}

\theoremstyle{definition}
\newtheorem{theorem}{Theorem}[section]
\newtheorem{proposition}{Proposition}[section]

\newtheorem{corollary}{Corollary}[section]
\theoremstyle{definition}
\newtheorem{definition}{Definition}[section]

\renewcommand{\l}{\Lambda}
\renewcommand{\H}{{\cal H}}
\newcommand{\bs}{\boldsymbol}
\newcommand{\p}{\hat{p}}
\newcommand{\<}{\langle}
\newcommand{\be}{\begin{equation}}
\renewcommand{\>}{\rangle}
\newcommand{\s}{\sigma}
\newcommand{\sm}{\mathsf{s}}

\renewcommand{\a}{\bs{\p}j_{3}[\sm j]}
\newcommand{\apr}{\bs{\p'}j'_{3}[\sm' j']}
\newcommand{\am}{\bs{\p}j_{3}[\sm j]^{-}}
\newcommand{\ap}{\bs{\p}j_{3}[\sm j]^{+}}
\newcommand{\amp}{\bs{\p}j_{3}[\sm j]^{\mp}}
\newcommand{\amd}{\bs{\p}j_{3}[\sm_{R} j_{R}]^{-}}
\newcommand{\amr}{\bs{\p}j_{3}[\sm j_{R}]^{-}}

\newcommand{\Bp}{\tilde{\cal S}\cap{\cal H}_{+}^{2}}
\newcommand{\Bm}{\tilde{\cal S}\cap{\cal H}_{-}^{2}}

\newcommand{\Bmp}{\tilde{\cal S}\cap{\cal H}_{\mp}^{2}}

\begin{document}
\title{Relativistic Gamow Vectors I\\
Derivation from Poles of the 
$S$-Matrix}
\author{A.~Bohm\footnote{bohm@physics.utexas.edu} 
\qquad H.~Kaldass\footnote{hani@physics.utexas.edu}
\qquad S.~Wickramasekara\footnote{sujeewa@physics.utexas.edu}  
\\ Physics Department\\ The University of Texas at Austin\\
Austin, Texas 78712} 
\date{}
\maketitle
\begin{abstract}
A state vector description for relativistic resonances  
is derived from the first order pole of the $j$-th partial $S$-matrix
at the invariant square mass value $\sm_R=(m-i\Gamma/2)^2$ in
the second sheet of the Riemann energy surface. To 
associate a ket, called Gamow vector, to the pole, we use
the generalized eigenvectors of the four-velocity operators in
place of the customary momentum eigenkets of Wigner,
and we replace the conventional Hilbert space assumptions
for the in- and out-scattering states with the new
hypothesis that in- and out-states are described by two different
Hardy spaces with complementary analyticity properties. 
The Gamow vectors have the following properties:\\
-They are simultaneous generalized eigenvectors of the four
velocity operators with real eigenvalues and of the self-adjoint 
invariant mass  
operator $M=(P_\mu P^\mu)^{1/2}$ with complex eigenvalue
$\sqrt{\sm_R}$.\\
- They have  
a Breit-Wigner distribution in the invariant square 
mass variable $\sm$ and lead to an exactly exponential law
for the decay rates and probabilities.
\end{abstract}
\maketitle

\section{Introduction}
This paper is the first of two papers which give
the theoretical and mathematical foundations and the detailed
derivations of certain  
results that were used in a preceding paper on the mass and width of
relativistic unstable particles~\cite{z-boson}. In particular, it
gives the
construction of relativistic Gamow vectors which were used
in~\cite{z-boson} to describe the states of a relativistic
resonance. This 
construction is done in complete analogy to the construction of 
the non-relativistic Gamow vectors 
in the Rigged Hilbert Space (RHS) formulation of time asymmetric quantum
mechanics~\cite{aB97,PRA}. We shall briefly review some of the
key properties of the non-relativistic theory in
the next section. However, the relativistic theory does not require
the non-relativistic theory as a backdrop and can be developed
starting with the representation spaces of the Poincar\'e group.

The existence of a fundamental time asymmetry in quantum
physics has been noted in the literature, e.g.,~\cite{gellmann}
and~\cite{haag}, 
and the irreversible character of quantum mechanical decay is fairly
well known~\cite{cohen}. However an exact mathematical theory of decay
that accommodates this property,
especially in the 
relativistic domain, did not exist. Time asymmetric boundary conditions
were incorporated in scattering theory, often unwittingly, by the
heuristic 
Lippmann-Schwinger equations~\cite{lippmann} 
for the in- and out- plane wave states $|E,b^\pm\>$. These were
represented by Dirac
kets which were supplementary to 
an otherwise time symmetric theory in Hilbert space, where the
time evolution is necessarily reversible and given by a unitary group.
In contrast, the time asymmetric quantum mechanics in Rigged Hilbert
Spaces~\cite{aB97, PRA} provides a precise definition of the
Lippmann-Schwinger Dirac kets $|E,b^\pm\>$ as functionals in a pair of
RHS's of Hardy class and defines the
non-relativistic Gamow vectors using the analytic extension of the
Lippmann-Schwinger kets into the
complex energy half-planes.

The standard relativistic quantum
theory is based on the unitary
irreducible representations (UIR) of the Poincar\'e
group~\cite{wigner,weinberg} which describe stable elementary particles.
More complicated relativistic quantum systems are described by direct
sums (towers) or by direct products (combinations) of UIR's. Within this
frame of unitary representations of the relativistic space-time
symmetry group,  
resonances cannot be described as elementary autonomous physical 
systems.

Our construction of the relativistic Gamow vectors as vectors
associated to the $S$-matrix pole at $\sm=\sm_R$ will lead 
to irreducible {\em semigroup} representations of causal Poincar\'e
transformations. Like Wigner's unitary group representations $(j,m^{2})$ of
space-time transformations, these semigroup representations are
characterized by two numbers 
$(j,\sm_{R})$, the spin $j$ and the invariant mass squared
$\sm_{R}=(M_R-i\Gamma_R/2)^{2}$.
The semigroup representations describe 
relativistic quasistable particles by relativistic Gamow vectors. Further, 
the transformation properties of these state vectors under Poincar\'e
transformations  
will show that their time evolution is exponential, and  
that the coefficient of this exponential time evolution is 
{\em precisely} $\gamma\Gamma_R/2$ ($\Gamma_R/2$ in the rest frame)
from which we expect that the lifetime is $\tau=\hbar/\Gamma_R$.
Using Gamow vectors for the $Z$-boson 
will thus define its width as $\Gamma_R=-2{\rm Im}\sqrt{\sm_R}$ 
and remove the ambiguity in the definition of
its mass and width~\cite{tR98,particledata}. 

The parameters
$(j,\sm_R)$ that characterize the semigroup representations should
be related to a definition of the relativistic resonance.
We therefore
start with the most widely accepted definition
of a resonance as  the first order pole of the partial $S$-matrix
$S_j(\sm)$ with angular momentum $j$ (perhaps spin-parity $j^\pi$)  
located at $\sm=\sm_R$ (in the second sheet).
We then analytically extend the relativistic Lippmann-Schwinger kets,
$|j,\sm,b^\pm\>$, from the ``physical'' values $\sm_0\leq\sm<\infty$
into the complex plane and define the relativistic Gamow kets
$|[j,\sm_R],b^-\>$  by the integral around the resonance pole 
at $\sm=\sm_R$.
These relativistic  Gamow kets are generalized eigenvectors
of the self-adjoint mass operator $\left(P_\mu
P^\mu\right)^{1/2}$ with complex 
eigenvalue $\sqrt{\sm_R}$. They provide the state vector description
of the physical entity described by the pole term  in the
relativistic $S$-matrix, i.e., the relativistic Breit-Wigner amplitude.  

Phenomenologically one often takes the point of view that resonances
or decaying particles
are autonomous quantum physical entities characterized by two real
values, either the resonance mass $m$ and resonance width 
$\Gamma$~\cite{particledata},
or the mass $m$ and lifetime $\tau$. One always assumed the relation 
$\tau=\hbar/\Gamma$, though this could be
established only in the Weisskopf-Wigner approximation and 
even the validity of the exponential law was questioned.

Stability or the value of the
lifetime does not appear to be a criterion for elementarity of a
relativistic particle. 
Therefore, a mathematical framework capable of both uniting the notions of
Breit-Wigner resonance and exponentially decaying particle and  
characterizing stable and quasi-stable states on the same footing has a strong
theoretical appeal.

Such a characterization has been accomplished for the non-relativistic
case~\cite{PRA}, 
where a
decaying state has been described by a generalized eigenvector
of the (self adjoint, semi-bounded) Hamiltonian 
with a complex eigenvalue 
$z_{R}=E_{R}-i\Gamma_R/2$ in much the same way as stable
particles are described by eigenvectors of the
Hamiltonian with real eigenvalues. 
The analogous case
for the relativistic theory will be discussed in this paper.

\section{Key Features of Non-relativistic Gamow\\ Vectors}

In the standard Hilbert space formulation of quantum mechanics, such
vectors with complex eigenvalues 
do not exist, and for their formulation in the non-relativistic case
one had to go to the Rigged Hilbert Space 
description of quantum mechanics. This was 
not a revolutionary step  since
the mathematical definition of Dirac kets already required the Rigged
Hilbert Space.
The generalized eigenvectors with complex eigenvalues,
$\psi^G\equiv |E_R-i\Gamma_R/2^{-}\>\sqrt{2\pi\Gamma}$, which we 
called Gamow vectors because of their property \eqref{evolutionplus}
below, were then obtain 
from the pole term of the $S$-matrix.
The real parameters $E_{R}$ and $\Gamma_R$
are respectively interpreted as resonance energy and resonance width,
for reasons that will become clear from their property  
\eqref{evolutionplus} below.

Like Dirac kets,
Gamow vectors are functionals of a Rigged Hilbert Space (Appendix A):
\begin{equation}
\label{triplet}
\Phi_{+}\subset \H \subset\Phi_{+}^{\times}\,:\quad \psi^{G}\in
\Phi_{+}^{\times}
\end{equation}
and the mathematical meaning of the eigenvalue equation 
$H^{\times}|z_{R}^{-}\>=z_{R}\,|z_{R}^{-}\>$ is:
\begin{equation}
\label{definition}
\<H\psi|z_{R}^{-}\>\equiv
\<\psi|H^{\times}|z_{R}^{-}\>=z_{R}\,\<\psi|z_{R}^{-}\>\,
\text{ for all }\, \psi \in \Phi_{+}\, .
\end{equation}
The conjugate operator $H^{\times}$ of the Hamiltonian $H$ is defined, 
by the first equality in \eqref{definition}, as the extension of the
Hilbert space adjoint operator $H^{\dagger}$ of $H$ to the space
of functionals $\Phi_{+}^{\times}$
\footnote{For (essentially) self-adjoint $H$, $H^{\dagger}$ is
equal to (the closure of) $H$; but we shall use the definition
\eqref{definition} also for unitary operator $U$ where 
$U^{\times}$ is the extension of $U^{\dagger}$ and not 
$U$.}.

The non-relativistic Gamow vectors have the following properties:
\begin{enumerate}
\item
They have an asymmetric time evolution and obey an exponential
law:
\begin{eqnarray}
\label{evolutionplus}
&\psi^{G}(t)=e^{-iH^{\times}t}_{+}|E_{R}-i\Gamma/2^{-}\>
=e^{-iE_{R}t}e^{-\Gamma t/2}|E_{R}-i\Gamma/2^{-}\>,\nonumber\\ 
&\text{\emph{ only for }}t\geq 0 \, .
\end{eqnarray}
There is another Gamow vector ${\tilde \psi^{G}}=|E_{R}+i\Gamma/2^{+}\>
\in \Phi_{-}^{\times}$, and a semigroup $e^{-iH^{\times}t}_{-}$
for $t\leq 0$ in another Rigged Hilbert Space 
$\Phi_{-}\subset\H\subset \Phi_{-}^{\times}$ (with the same 
$\H$) with 
the asymmetric evolution
\begin{eqnarray}
\label{evolutionminus}
&{\tilde \psi^{G}}=e^{-iH^{\times}t}|E_{R}+i\Gamma/2^{+}\>
=e^{-iE_{R}t}e^{\Gamma t/2}|E_{R}+i\Gamma/2^{+}\>,\nonumber\\ 
&\text{\emph{ only for }}
t\leq 0 \, .
\end{eqnarray}
\item The $\psi^{G}\,({\tilde\psi^{G}})$
is derived as a functional associated to the resonance pole term 
at $z_{R}=E_{R}-i\Gamma/2$ (at $z_{R}^{*}=E_{R}+i\Gamma/2$)
 in the second sheet of the analytically continued $S$-matrix.
\item The Gamow vectors have a Breit-Wigner energy distribution
\begin{equation}
\label{breitwigner}
\<^{-}E|\psi^{G}\>=i\sqrt{\frac{\Gamma}{2\pi}}\,
\frac{1}{E-(E_{R}-i\Gamma/2)}\, , \, -\infty_{II}<E<\infty\, .
\end{equation}
where $-\infty_{II}$ means that it extends to $-\infty$
on the second sheet of the $S$-matrix (whereas the standard 
Breit-Wigner terminates at the threshold $E=0$).
\item The decay probability 
$P(t)=Tr(\Lambda_{\eta}|\psi^{G}(t)\>\<\psi^{G}(t)|)$
of $\psi^{G}(t),\, t\geq 0$, into the final non-interacting decay 
products of the channel described by $\Lambda_\eta$ 
can be calculated as a function of time. From this the
decay rate $R(t)=\frac{dP(t)}{dt}$ is obtained by differentiation
as $R(t)=\Gamma^{(\eta)}e^{-\Gamma t}$ where $\Gamma^{(\eta)}$
is the partial width for the decay channel $\eta$.
And it leads to an exact Golden rule for $\Gamma^{(\eta)}$
which in the Born approximation ($\psi^{G}\rightarrow
f^{D}$, an eigenvector of $H_{0}=H-V$; $\Gamma/E_{R}\rightarrow 0$;
$E_{R}\rightarrow E_{0}$) goes into Fermi's Golden rule No.$2$ of Dirac.
As a consequence of the exponential law for $R(t)$, the lifetime
of the state described by $\psi^{G}(t)$ is given precisely 
by $\tau=\hbar/\Gamma$.
\end{enumerate}

We want to generalize these non-relativistic Gamow vectors to the
relativistic case and construct representations of the Poincar\'e
group $\cal P$
that describe relativistic resonances and decaying states very much in the
same way as unitary irreducible representations (UIR) 
of $\cal P$ describe stable relativistic particles.
In order to obtain a state vector description for unstable particles,
one has to start from the space of decay products. 
For a relativistic unstable particle decaying
into a two-particle system with masses $m_{1}$, $m_{2}$, and spins
$s_{1}$, $s_{2}$, the (asymptotically free) decay product space is 
the direct product space of irreducible
representation spaces of the two particles $\H(m_i,s_i)$, $i=1,\,\,2$:
$\H_{12}=\H(m_1,s_1)\otimes\H(m_2,s_2)$. A decaying state vector
can then be obtained, in analogy to the non-relativistic case, by extending
the basis vectors of $\H_{12}$ to complex energy, in a manner which is 
consistent with the dynamical description of unstable particles as
poles of the analytically continued $S$-matrix. Since resonances
occur in one-particular partial-wave $j_R^\pi$, (which designates the spin
(total angular momentum in the rest frame) and parity of the unstable 
particle), we need to use
the basis vectors of $\H_{12}$ which 
are diagonal in total angular momentum in order to obtain the
relativistic Gamow vectors.
These angular momentum basis vectors are obtained by the reduction of
the direct product $\H(m_1,s_1)\otimes\H(m_2,s_2)$ into a continuous
direct sum of irreducible representation spaces 
\begin{equation}
\label{star}
\H_{12}=\sum_{j\eta}\int_{(m_1+m_2)^{2}}^{\infty}d\mu(\sm)\H^{\eta}_{n}
(\sm,j)\,,
\end{equation}
where $\sm$ is the invariant mass square (Mandelstam parameter)
for the two-particle system $\sm=(p_1+p_2)^{2}$, and $\eta$ and $n$
are degeneracy and particle species labels 
respectively~\cite{wightman,hani}. 
As discussed in Sections~\ref{setup} and~\ref{a} below, the partial
$S$-matrix is a function of $\sm$  which is analytic
except for poles and branch cuts. 
The complex poles at $\sm=\sm_R$ are the ones associated with the 
unstable particles, and to these values the analytic extension in 
the $\sm$-variable will be performed.

The extension of the invariant mass square $\sm$ 
to a complex value leads necessarily to complex momenta, since
$\sm=E^2-\bs{p}^2$. As Lorentz transformations intermingle energy and momenta,
this in general leads to complex momentum representations of the
Poincar\'e group.
To obtain a description of unstable particles with minimal
modifications to the stable particle case, and to restrict
the set of complicated complex momentum representations of 
${\cal P}$~\cite{cm},
we will consider representations of ${\cal P}$ where complexness
is due only to the complexness of the mass square $\sm$. 
These representations are ``minimally complex'', in the sense that,
while the invariant mass squared $\sm$ is complex, the $4$-velocities
$\p^\mu=p^\mu/\sqrt{\sm}$ remain real. 
This construction was motivated by a remark by D.~Zwanziger~\cite{zwanziger}
and is based on the fact that the velocity
kets provide as valid a basis for the UIR's of the
Poincar\'e group as Wigner's momentum
kets. Moreover, the $4$-velocity eigenvectors are often more useful for 
physical reasoning, because 
$4$-velocities seem to fulfill to a rather good approximation 
``velocity super-selection rules'' 
which the momenta do not \cite{velocityvectors}.

The reduction of $\H_{12}$ in the velocity-angular momentum 
basis $|\bs{\p}j_{3}[\sm j]\eta,n\>$ was performed in \cite{hani}
as a preparation for  this paper. 
The discussion there extends to an unstable particle 
decaying into more than two particles,
but the reduction of the decay product space into a direct sum 
of UIR spaces becomes  much more tedious than for the two-particle
case. Hence, in general, with the use of velocity-angular momentum kets,
the relativistic Gamow vectors will be defined not as
momentum eigenvectors but as $4$-velocity eigenvectors in the 
direct product space of UIR's for the decay products of the 
resonance $R$. This definition requires the extension of the invariant
square mass 
$\sm=(p_{1}+p_{2}+\cdots)^{2}$, 
where $p_{1},\, p_{2} \cdots$ are the momenta of the decay 
products of $R$, into the complex values. 
In analogy to the non-relativistic case, we define 
the resonances
$R$ by poles of the analytically continued relativistic
partial $S$-matrix with total angular momentum $j=j_{R}$ and 
pole position $\sm=\sm_{R}=(M_{R}-i\Gamma_{R}/2)^{2}$. 
As discussed in Section~\ref{introduction} 
below, there appear to be phenomenological
and theoretical arguments in favor of this definition. 
This means that the 
relativistic particle is characterized by the value of its spin
$j_{R}$ and its mass $w_{R}\equiv\sqrt{\sm_R}=M_{R}-i\Gamma_{R}/2$, where 
$\Gamma_{R}$
is zero for stable particles, but $w_{R}$ is complex if the particle 
is unstable. 
Just as the mass and spin of a stable particle permit its association
to an irreducible unitary representation of the Poincar\'e group
characterized by these values, it will be shown in a forthcoming
paper that the complex mass $w_{R}$ and spin
$j_R$ concatenate an unstable particle to an irreducible representation
of the Poincar\'e group, only these irreducible representations
are not unitary. In fact, it turns out
that they are irreducible representations of the causal Poincar\'e 
{\em semigroup}, defined as the semi-direct product of the 
group of homogeneous Lorentz transformations with the {\em semigroup}
of space-time translations into the forward light cone. The fundamental
mathematical object needed in constructing these semigroup
representations is the relativistic Gamow vector. The main technical
result of this paper is its derivation, while the discussion
of the semigroup representations is deferred to a later paper.

\section{Relativistic Gamow Vectors From Poles of the Relativistic
$\bs{S}$-matrix}

\subsection{Introduction}\label{introduction}

As discussed above, our aim is to
obtain the relativistic Gamow vectors from the pole term of
the relativistic $S$-matrix in complete analogy to the way the 
non-relativistic Gamow vectors were obtained 
\cite{our}. For the sake of definiteness, we discuss here the case
of resonance formation in an elastic scattering process
$a\,b\rightarrow R \rightarrow a\,b$ for which one 
may consider
$\pi^+\pi^-\rightarrow\rho\rightarrow\pi^+\pi^-$~\cite{zimmermann} or
$e^+ e^-\rightarrow Z\rightarrow e^+ e^-$~\cite{tR98} as examples.

In the absence of
a vector space description of a resonance, we shall also in 
the relativistic theory 
define the unstable particle by the pole
of the analytically continued partial $S$-matrix 
with angular momentum $j=j_{R}$ at the value
$\sm=\sm_{R}\equiv(M_{R}-~\frac{i}{2}\Gamma_{R})^{2}$ 
of the invariant mass
square variable (Mandelstam variable) $\sm$ \cite{eden}.
This means that the mass $M_{R}$, width $\Gamma_R$ (which will be shown
to be connected to the lifetime by $\tau_R=\hbar /\Gamma_{R}$), and  
spin $j_{R}$ are the intrinsic properties that define a quasistable
relativistic particle
\footnote{The more conventional definition, used in~\cite{particledata},
of resonance mass and width is $M_\rho$, $\Gamma_\rho$, which are defined
in terms of $M_R$, $\Gamma_R$ by
\begin{equation}
\tag{\ref{idealbreitwigner}$\rm{b}$}
\sm_{R}\equiv M_{\rho}^{2}-iM_{\rho}\Gamma_{\rho}
=M_{R}^{2}\left(1-\frac{1}{4}\left(\frac{\Gamma_{R}}{M_{R}}\right)^{2}
\right)
-iM_{R}\Gamma_{R}
\end{equation}
or $M_{\rho}=M_{R}\sqrt{1-\frac{1}{4}
\left(\frac{\Gamma_{R}}{M_{R}}\right)^{2}}$ 
and $\Gamma_{\rho}
=\Gamma_{R}\left(1-\frac{1}{4}\left(\frac{\Gamma_{R}}{M_{R}}\right)^{2}
\right)^{-1/2}$. 
$M_\rho^{2}$ is the peak position of the relativistic Breit-Wigner
probability $|a_{j_{3}}(\sm)|^{2}$.
The exponential time evolution of the Gamow vectors which will be 
derived in the forthcoming sequel to this paper, shows that $\Gamma_R$,
and not $\Gamma_\rho$, is the inverse lifetime. Since only $\Gamma_R$
(not $\Gamma_Z$ or $\Gamma_\rho$) of the parameterization 
\eqref{idealbreitwigner} precisely fulfills the identity 
$\Gamma_R=\hbar/\tau$, we call $\Gamma_R$ the ``width'' of the
relativistic resonance.
For the $\rho$-meson 
$\left(\frac{\Gamma_{R}}{M_{R}}\right)^{2}\thickapprox 0.03$
and for the $Z$-boson 
$\left(\frac{\Gamma_R}{M_R}\right)^{2}\approx 7\times 10^{-4}$.
Usually the difference between $M_\rho$ and $M_R$
is an order of magnitude smaller and/or well within the
experimental error for the experimental mass; however for the 
$\rho$-meson and in particular for the $Z$-boson this difference is 
significant, see~\cite{cern}.}.
As will be shown below in Section~\ref{continuation}, the 
$j$-th partial $S$-matrix can be separated into a background term
and a pole term which is given
by the relativistic Breit-Wigner amplitude
\begin{equation}
\label{idealbreitwigner}
a_{j_{R}}=\frac{f_{n}\Gamma_{R}/2}{\sm-(M_{R}-\frac{i}{2}\Gamma_{R})^{2}}
\quad\text{ with }
-\infty_{II}\leq \sm \leq \infty \, , 
\end{equation}
where the negative values of $\sm$ are on the second sheet.

\addtocounter{footnote}{-1}
However we want to mention that the $S$-matrix pole definition 
with the parameterization by $M_R$, $\Gamma_R$
is neither the only nor the universally accepted definition of the
parameters mass $m$ and width $\Gamma$ of a relativistic resonance. 
The particle data table uses $M_\rho$
and $\Gamma_\rho$ as mass and width~\footnotemark.
And for precision data as in the line shape analysis for the 
$Z$-boson~\cite{tR98}, one prefers
an energy dependent width
replacing $M_{\rho}\Gamma_{\rho}\rightarrow \sqrt{\sm}\,\Gamma(\sm)$,
cf.\ equation $(35.53)$ of \cite{particledata}. 
This gives still another parameterization which in the case of the
$Z$-boson is called $M_Z$, $\Gamma_Z$~\cite{tR98}.
In the line-shape analysis for the $Z$-boson,
these three parameterizations lead to three different values for 
the $Z$-boson mass and width~\cite{cern} which fit the experimental
data equally well.
Therefore the line-shape data do not provide a phenomenological
clue as to how the fundamental parameters $m$, 
$\Gamma$ of the $Z$-boson should be defined~\cite{cern}.
In the case of the $\rho$-meson,
the definition of
the resonance parameters given by \eqref{idealbreitwigner}
leads to better,
process-independent fits of the parameters \cite{lC98}. 
Thus, there seems 
to be a slight phenomenological preference for 
the $S$-matrix pole definition~\eqref{idealbreitwigner}
of a relativistic resonance.
The parameterizations in terms of 
$M_{R}$, $\Gamma_{R}$ or $M_\rho$ $\Gamma_\rho$ of the
$S$-matrix pole position $\sm_R$ are phenomenologically equivalent. 
One of the results of our theory
will be that only $\Gamma_R$ and not $\Gamma_\rho$ (or $\Gamma_Z$)
will be connected to the lifetime by $\Gamma_R=\hbar/\tau_R$,
which is the theoretical argument in favor of the parameterization
$\sm_R=(M_R-i\Gamma_R/2)^{2}$.

In order to make the analytic continuation in the partial $S$-matrix
with angular momentum $j$, we need the angular momentum basis vectors
for the scattering states.
Thus we replace the non-relativistic angular momentum
basis vectors in the derivation for the non-relativistic Gamow vectors 
with:
\begin{align}
\label{replacefree}
|\,E\,\rangle&\equiv |\,Ell_{3}\eta\,\rangle  
\rightarrow |\,\bs{\p}j_{3}[w=\sqrt{\sm},j]\,\rangle \\
\tag{\ref{replacefree}$^{\pm}$}\label{replaceint}
|\,E^{\pm}\,\rangle&\equiv |\,Ell_{3}\eta^{\pm}\,\rangle\rightarrow
|\,\bs{\p}j_{3}[w=\sqrt{\sm},j]^{\,\pm}\,\rangle\, .
\end{align}
The two sets of bases are related by the Moeller wave operators
$$
|\,\bs{\p}j_{3}[w=\sqrt{\sm},j]^{\,\pm}\,\rangle
=\Omega^{\pm}|\,\bs{\p}j_{3}[w=\sqrt{\sm},j]\,\rangle\,.
$$
We shall analytically extend the Dirac kets~\eqref{replaceint}
in the variable $\sm$ from the
physical values to the complex values on the Riemann
surface of the $S$-matrix.
The analytic extension of $|\bs{\p},j_3,[\sm,j]^-\>$ 
at the pole position $\sm_R$ yields
the relativistic Gamow vector.
\addtocounter{footnote}{1}

\subsection{The $\bs{S}$-matrix~$^{3}$}\label{setup}
\footnotetext{Here we closely follow chapter $3$ of 
\cite{weinberg} in order to both display the analogy (and comparability)
and expose the differences 
between our development and the standard views in relativistic
quantum theory. Our notation transcribes into that of~\cite{weinberg}
as $\{\phi^{in/out}, \psi^{out}\}\rightarrow\Phi_{g}$ and 
$\{ \phi^{+},\psi^{-}\}\rightarrow \Psi^{\pm}_{g}$.
In~\cite{weinberg}, the multi-particle basis vectors
are also denoted by $\Psi^{\pm}_{\alpha}$ where $\alpha=\{p_{1}\s_{1}
n_{1},p_{2}\s_{2}n_{2},\cdots\}$, $\s$ is the third component
of the spin, and $n$ is the species label.}\label{f}

In a scattering experiment, the experimentalist prepares 
an initial state
$\phi^{in}$~\footnote{In realistic experiments the states are not pure but
mixtures $W^{in}=\sum w_{\alpha}|\phi^{in}_{\alpha}\>\<\phi_{\alpha}^{in}|$
and the observables are not given by projection operators
but by
$\Lambda^{out}=\sum \lambda(\beta)|\psi_{\beta}^{out}\>\<
\psi_{\beta}^{out}|$}, 
describing the non-interacting projectile and target beams,
at $t\rightarrow -\infty$. Then, later, at $t'\rightarrow +\infty$
the experimentalist measures or registers an observable~\footnotemark[4]
 $|\psi^{out}\>\<\psi^{out}|$. It is assumed that the time translation
generator $H$ can be divided into two terms, the ``free-particle''
Hamiltonian $K$($=P_{1}^{0}+P_{2}^{0}$ at rest) 
and an interaction part $V$:
$$H=K+V\, ,$$
where the split of $H$ into $K$ and $V$ will be different if
different in- and out-particles are involved.

The state vectors $\phi^{in}(t)=e^{-iKt}\phi^{in}$ and the observable
vectors $\psi^{out}(t')=e^{-iKt'}\psi^{out}$ evolve in time according 
to the free Hamiltonian $K$.
When the beams reach the
interaction region, the free in-state vector $\phi^{in}$ turns into the exact
state vector $\phi^{+}$ whose time evolution is given by the exact
Hamiltonian $H=K+V$:
\begin{equation}
\label{timeevolution}
\Omega^{+}\phi^{in}(t)\equiv\phi^{+}(t)=e^{-iHt}\phi^{+}=
\Omega^{-}\phi^{out}(t)\, .
\end{equation}
Here $t$ is the proper time in the center-of-mass of the projectile and
target. This vector $\phi^{+}$ 
leaves the interaction region and becomes the well 
determined state $\phi^{out}$. The state vector $\phi^{out}$ is 
determined from $\phi^{in}$ by the dynamics of the scattering process:
\begin{equation}
\label{smatrix}
\phi^{out}=S\phi^{in},\quad S=\Omega^{-\dagger}\Omega^{+}
\end{equation}
The state $\phi^{in}$ and thus $\phi^{+}$ and also $\phi^{out}$ 
are determined by the preparation apparatus (the accelerator). 

A scattering experiment
consists of a preparation apparatus and a registration apparatus (detector).

The registration apparatus registers an observable $|\psi^{out}\rangle
\langle\psi^{out}|$ outside the interaction region. This observable
vector $\psi^{out}$ comes
from a vector
\begin{equation}
\label{psiminus}
\psi^{-}=\Omega^{-}\psi^{out}
\end{equation}
in the interaction region. The observable $\psi^{out}$ is of 
course not the same as the state $\phi^{out}$, since $\phi^{out}$, and 
thus $\phi^{+}$ and $\phi^{in}$, is defined by the 
accelerator and $\psi^{out}$, and thus $\psi^{-}$, is defined by the 
detector. Hence he set of vectors $\psi^{-}$ may be, and in our case
is, distinct from the set of vectors $\phi^{+}$.

The probability amplitude to register the observable $|\psi^{out}\>
\<\psi^{out}|$ in the state $\phi^{out}$, which--by
simple calculations using 
\eqref{timeevolution} and \eqref{smatrix}--is the same as the
probability amplitude for the observable $|\psi^{-}\>\<\psi^{-}|$ in the
state $\phi^{+}$, is the scalar product 
$$
(\psi^{out},\phi^{out})=(\psi^{out},S\phi^{in})
=(\Omega^{-}\psi^{out},\Omega^{+}\phi^{in})=(\psi^{-},\phi^{+})\, .
$$
The fundamental probabilities of quantum theory (Born probability)
are the probabilities for the observables $|\psi^{-}\>\<\psi^{-}|$ 
in the state
$|\phi^{+}\>\<\phi^{+}|$ (or for the observable vector $\psi^{-}$
in the state vector $\phi^{+}$) given by
$$Tr(|\psi^{-}\>\<\psi^{-}\,|\,\phi^{+}\>\<\phi^{+}|)=|(\psi^{-},
\phi^{+})|^{2}\, .$$

The vectors $\phi_{\alpha}^{+}$ are called the in-states and the vectors
$\psi_{\beta}^{-}$ are called the out-states and the array
of complex amplitudes $(\psi_{\beta}^{-},\phi_{\alpha}^{+})$ is 
called the $S$-matrix. The labels $\alpha$ and $\beta$ stand for a whole
collection of {\it{discrete}} quantum numbers. The $S$-matrix
is also defined when $\alpha$ and/or $\beta$ are continuous labels
(of basis vectors for the space of the $\{\psi^{-}\}$ and the space
of the $\{\phi^{+}\}$) only then the $S$-matrix does not
represent probability but a probability density. 
``It should be stressed that in-states $\phi^{+}$ and out-states
$\psi^{-}$ 
do not inhabit two different Hilbert spaces''~\cite{weinberg}
but they do inhabit 
two {\it different} (dense) subspaces of the same Hilbert space
$\cal H$; these two dense subspaces 
we call $\Phi_-\equiv\{\phi^+\}$ and $\Phi_+\equiv\{\psi^-\}$.
This is the only new hypothesis by which our time asymmetric 
quantum theory in Rigged Hilbert Space differs from the standard
Hilbert space quantum theory. Thus the Rigged Hilbert Space theory
distinguishes meticulously between prepared states (in-states) 
$\{\phi^+\}$ and observables (out-states) $\{\psi^-\}$ 
by means of two different Rigged Hilbert Spaces of Hardy class:
\begin{equation}
\label{triplets}
\begin{split}
\!\!\!\!\!\!\!\!
\phi^{+}\in \Phi_{-}\subset {\H}\subset \Phi_{-}^{\times}\quad
\parbox[t]{3in}{ for the prepared in-states defined by
the preparation apparatus (accelerator)}\\
\!\!\!\!\!\!\!\!
\psi^{-}\in \Phi_{+}\subset {\H}\subset \Phi_{+}^{\times}\quad
\parbox[t]{3in}{ for the registered out-states (observables) defined
by the detector}
\end{split}
\end{equation}
In contrast, the conventional scattering theory assumes
$\Phi_-=\Phi_+(={\cal H})$.
As a consequence of this new hypothesis,
an arbitrary in-state $\phi^{+}\in \Phi_{-}$ {\em cannot}
be expanded as a sum of out-states $\psi_{\alpha}^{-}\in \Phi_{+}$.
However (for instance in the non-relativistic theory), 
any in-state $\phi^{+}$ can be expanded with respect to a 
generalized basis system $|E^{-}\>\in\Phi_{+}^{\times}$ of the 
out-states $\psi^{-}$ ($\psi^{-}=\int dE |E^{-}\>\<^{-}E\,|\,\psi^{-}\>$)
using the $S$-matrix :
$$
\phi^{+}=\int_{0}^{\infty}dE|E^{-}\>\<^{-}E\,|\,\phi^{+}\>
=\int_{0}^{\infty}dE |E^{-}\>S(E+i0)\<^{+}E\,|\,\phi^{+}\>\, .$$


In the non-relativistic theory the
boundary conditions required of  the spaces $\Phi_{-}$ and
$\Phi_{+}$ lead to distinct analyticity properties
for the energy wave functions $\<^{+}E\,|\,\phi^{+}\>$ 
of $\phi^{+}\in \Phi_{-}$ and the energy wave functions
$\<^{-}E\,|\,\psi^{-}\>$ of $\psi^{-}\in\Phi_{+}$ \cite{antoniou}.
To wit
\begin{subequations}
\label{hardy}
\begin{gather}
\label{hardy+}
\psi^{-}\in \Phi_{+}\text{  if and only if  }
\langle ^{-}E|\psi^{-}\rangle=\langle E|\psi^{out}\rangle
\in {\cal S}\cap {\cal H}_{+}^{2}|_{\mathbb{R}_{+}}\\
\label{hardy-}
\phi^{+}\in \Phi_{-}\text{  if and only if  }
\langle ^{+}E|\phi^{+}\rangle=\langle E|\phi^{in}\rangle
\in {\cal S}\cap {\cal H}_{-}^{2}|_{\mathbb{R}_{+}}
\end{gather}
\end{subequations}
where: $\cal S$ denotes the Schwartz space; $\H^{2}_{\pm}$, the Hardy
class functions (Appendix~\ref{h}) 
which are boundary values of functions that are analytic
in the open $\begin{array}{cc}
\text{upper}\\
\text{lower}
\end{array}$ half plane of the complex energy plane; 
and $|_{{\mathbb R}_+}$, the restriction to the physical
values of the energy, ${\mathbb R}_{+}=[0,\infty)$. 
As the lower half complex energy plane we take the second 
Riemann sheet of the $S$-matrix reached from the physical values
$\<^{+}E+i0\,|\,\phi^{+}\>$, $E\in \mathbb{R}_{+}$ by burrowing
down through the cut along the positive real axis $\mathbb{R}_{+}$. As the
upper half complex energy plane we take the second Riemann sheet of
the $S$-matrix reached from the values \hfill\\
$\<^{-}E-i0\,|\,\psi^{-}\>$,
$E\in\mathbb{R}_{+}$ on the physical sheet by burrowing up through the cut
along $\mathbb{R}_{+}$.
In the non-relativistic case \eqref{hardy-} and \eqref{hardy+} 
could be obtained
from a mathematical formulation of causality (using
the Paley-Wiener theorem)~\cite{antoniou}.
An important consequence of the distinction between 
$\Phi_+$ and $\Phi_-$ obtained in \eqref{hardy} is that the time 
evolution operator $U(t)=e^{iHt}$, which is represented by 
a group on the Hilbert space $\H=L^{2}({\mathbb R}_{+})$, splits
into two semigroups when restricted to $\Phi_\pm$~\cite{our}.
This means that $U(t)|_{\Phi_{+}}$ is a continuous operator on $\Phi_+$
only for $t\geq 0$ and $U(t)|_{\Phi_{-}}$ is a continuous operator on
$\Phi_{-}$ only for $t\leq 0$.
It is this semigroup property that is at the origin
of the time asymmetric evolution equations \eqref{evolutionplus}
and \eqref{evolutionminus}.

In the relativistic theory, in particular for the 
elastic scattering of two particles, the Hilbert
space $\H$ of~\eqref{triplets} is the direct integral~\eqref{star} of 
irreducible representation spaces 
$\H_{n}^{\eta}(\sm,j)$
labeled by the Mandelstam 
variable $\sm=(p_1+p_2)^{2}$ and the total angular momentum
$j$ of the two particles. 
For the measure $d\mu(\sm)$ in \eqref{star} we take
the Lebesgue measure $d\sm$. 
In what follows, we will restrict our discussion to the elastic
scattering of two spinless particles with the same mass $m$.
We use as a basis for $\Phi_\pm$ the angular momentum velocity basis vectors,
\{$|\amp\>$\}~\cite{hani}. With these basis
vectors, the Dirac basis vector expansion for an in-state (of 
two equal mass spinless particles) $\phi^+\in\Phi_-$ is:
\begin{equation}
\label{t1}
\phi^+=\sum_{jj_{3}}\int_{4m^{2}}^{\infty}d\sm
\int\frac{d^{3}\p}{2\p^0}|\ap\>\<\ap|\phi^+\>\,,
\end{equation}
and for an out-observable $\psi^-\in\Phi_+$ is:
\begin{equation}
\label{t2}
\psi^-=\sum_{jj_3}\int_{4m^2}^\infty d\sm\int 
\frac{d^{3}\p}{2\p^0}|\am\>\<\am|\psi^-\>\,.
\end{equation}
In~\eqref{t1} and~\eqref{t2}, $\<\ap|\phi^+\>$ and $\<\am|\psi^-\>$
are the wave functions of $\phi^+$ and $\psi^-$ along the velocity
basis vectors. We shall make the hypothesis that these wave functions have the
same analyticity properties in the invariant mass squared 
$\sm=(E^{\rm{cm}})^{2}$ as the energy wave functions
in~\eqref{hardy}. 
However, in the relativistic case, due to the 
mathematical requirement of the invariance of the subspaces $\Phi_\mp$
under the action of the generators of the Poincar\'e group, a 
closed subspace $\tilde{\cal S}$ of the Schwartz space ${\cal S}$ 
of~\eqref{hardy}
has to be considered. The subspace $\tilde{\cal S}$, 
constructed in~\cite{sujeewa}, 
is the space of Schwartz functions which vanish at
zero faster than any polynomial.
This requirement also assures that the zero mass states do not
contribute to the Gamow vector (see~\eqref{rgv}). This avoids
the difficulty that the four velocity operators, which is centrally
significant to our construction of Gamow vectors, cannot be 
meaningfully defined in the zero-mass case in any obvious way.
The features of the space $\tilde{\cal S}$
which are needed for the construction of the relativistic Gamow
vectors are as follows~\cite{sujeewa}: 
\begin{proposition}\label{Property2.1}
The triplets
\begin{equation}
\label{striplet}
\Bmp|_{{\mathbb R}_{{\sm}_0}}
\subset 
L^{2}({\mathbb R}_{{\mathsf s}_0})
\subset
\left(\Bmp|_{{\mathbb R}_{{\sm}_0}}\right)^{\times}
\end{equation}
form a pair of Rigged Hilbert Spaces.
\end{proposition}
\noindent In~\eqref{striplet}, $\mathbb{R}_{\sm_{0}}$ is the set of
physical values of the Mandelstam variable $\sm$ for the scattering
process,
$\mathbb{R}_{\sm_0}=[(m_1+m_2)^2,\infty)$.
\begin{proposition}\label{Property2.2}
The space ${\tilde{\cal S}}$ is endowed with a nuclear Fr\'echet topology
such that multiplication by ${\sm}^{\frac{n}{2}}$,
$$
\sm^{\frac{n}{2}}\,:\,\,\tilde{{\cal S}}\cap{\cal H}_{\pm}^{2}
\rightarrow \tilde{{\cal S}}\cap{\cal H}_{\pm}^{2}\,,\quad
n=1,2,3,\cdots
$$
is a continuous linear operator in the topology of $\tilde{\cal S}$ .
\end{proposition}
Thus the relativistic characterization of $\Phi_\pm$ 
analogous to~\eqref{hardy+} and~\eqref{hardy-} is:
\begin{subequations}
\label{t}
\begin{equation}
\label{t3}
\psi^-\in\Phi_+\quad\text{if and only if  }
\<\am|\psi^-\>\in\Bp|_{\mathbb{R}_{\sm_0}}
\times{\cal S}(\mathbb{R}^3)
\end{equation}
\begin{equation}
\label{t4}
\phi^+\in\Phi_-\quad\text{if and only if  }
\<\ap|\phi^+\>\in\Bm|_{\mathbb{R}_{\sm_0}}
\times{\cal S}(\mathbb{R}^{3})\,;
\end{equation}
\end{subequations}
where ${\mathbb R}^{3}$ is the space of components of the
$4$-velocity
and the Hilbert space $\H$ of~\eqref{triplets} is realized by the
function space
\begin{equation}
\label{hilbert}
L^{2}({\mathbb R}_{\sm_0},d\sm)\times L^{2}\left({\mathbb R}^3,
\frac{d^{3}\p}{2\p^0}\right)\,.
\end{equation}
In~\eqref{t} as in~\eqref{hardy}, 
$\H_{+}^{2}$ means the functions of 
Hardy class analytic in the upper half of the second sheet
of the $\sm$-plane
and $\H_{-}^{2}$ means the functions of 
Hardy class analytic in its lower half.
Specifically, the physical values 
$\<^{+}\sm-i0\,|\,\phi^{+}\>$ are the  
boundary values of functions analytic in the lower half of the second
sheet and the $\<^{-}\sm+i0\,|\,\psi^{-}\>$ are the boundary values
of functions analytic in the upper half of the second sheet. These
analyticity properties on the second sheet of the complex
$\sm$-Riemann surface will turn out to be important because the (pairs
of) resonance poles of the $S$-matrix are located on the second 
Riemann sheet.

By virtue of Proposition~\ref{Property2.2}, 
the total momentum operators $P_\mu=P_{1\mu}+P_{2\mu}$ and the invariant
mass square operator $M^{2}=P_\mu P^\mu$ are 
$\tau_{\Phi_{\pm}}$-continuous operators; 
hence their conjugates~\footnote{defined by the 
first equality in~\eqref{reals}}, $P_{\mu}^{\times}$
and $M^{2^{\times}}$, are well defined on $\Phi_{\pm}^\times$.
This can be seen by considering the realization, for instance, of the vectors
$P_\mu \psi^-$ and $M^{2}\psi^-$:
\begin{subequations}
\label{reals}
\label{realization}
\begin{align}
\label{real-a}
\<P_\mu \psi^-|\am\>&=\<\psi^-|P_\mu^\times|\am\>
=\sqrt{\sm}\p_\mu\<\psi^-|\am\>\,,\\
\label{real-b}
\<M^2 \psi^-|\am\>&=\<\psi^-|M^{2^\times}|\am\>=\sm\<\psi^-|\am\>\,.
\end{align}
\end{subequations}
According to Proposition~\ref{Property2.2} and the definition of the 
wave functions $\<\psi^-|\am\>$ given in~\eqref{t3}, the multiplication 
operators by $\sqrt{\sm}\p_{\mu}$ and $\sm$ which appear in 
the right hand side of~\eqref{real-a} and~\eqref{real-b} are
$\tau_{\Phi_+}$-continuous. Consequently, $P_\mu$ and $M^{2}$
are $\tau_{\Phi_+}$-continuous operators, and the conjugate
operators $M^{2^{\times}}$ and $P_{\mu}^{\times}$ that appear
in~\eqref{realization} are everywhere defined, weak$^*$-continuous
operators on $\Phi_+^\times$.  
Hence,~\eqref{real-a} and~\eqref{real-b} define the functionals 
$|\am\>$ as generalized eigenvectors of $P_\mu$ and 
$M^{2}$. The same discussion applies for the space $\Phi_-$. Summarizing
\begin{equation}
\label{summary1}
P_\mu\,:\,\Phi_{\pm}\rightarrow\Phi_{\pm}
\quad\text{ is }\ 
\tau_{\Phi_{\pm}}\text{-continuous}
\end{equation}
and
\begin{subequations}
\label{summary2}
\begin{equation}
\label{summary2.1}
P_\mu^\times|\amp\>=\sqrt{\sm}\p_\mu |\amp\>\,,
\end{equation}
\begin{equation}
\label{summary2.2}
M^{2^\times}|\amp\>=\sm|\amp\>\,.
\end{equation}
\end{subequations}
We can re-express the generalized eigenvalues of the momentum operator
in terms of the three velocity $\bs{v}$ by noting that
$\bs{\p}=\gamma\bs{v}=\frac{\bs{v}}{\sqrt{1-\bs{v}^{2}}}$,
and $1+\bs{\p}^{2}=\frac{1}{1-\bs{v}^{2}}=\gamma^{2}$.
Hence, the eigenvalues in~\eqref{summary2.1} can be rewritten
as
\begin{equation}
\label{2.15.5}
\begin{split}
H^\times|\amp\>&=\gamma\sqrt{\sm}|\amp\>\,,\\
\bs{P}^\times|\amp\>&=\gamma\sqrt{\sm}\bs{v}|\amp\>\,.
\end{split}
\end{equation}

For the branch of $\sqrt{\sm}$ in~\eqref{realization}, \eqref{summary2}
and \eqref{2.15.5}, we choose
\begin{equation}
\label{branch}
-\pi\leq \rm{Arg}\,\sm<\pi\,.
\end{equation}
This choice of branch, even though irrelevant for the physical values
of $\sm$, will be needed since we will analytically continue
the kets $|\amp\>$ to the unphysical second Riemann sheet as described in
Section~\ref{continuation}.

We shall now consider the $S$-matrix element
\begin{multline}
\label{smatrixelement}
(\psi^{out},S\phi^{in})=(\psi^{-},\phi^{+})\\
=\sum_{jj_{3}}\int \frac{d^{3}\hat{p}}{2\hat{E}}d\sm
\sum_{j'j_{3}'}\int \frac{d^{3}\hat{p}'}{2\hat{E}'}d\sm'
\langle\,\psi^{-}\,|\,\am\rangle\\
\langle\,\a\,|\,
S\,|\,\apr\,\rangle
\langle^{+}\apr\,|\,\phi^{+}\,\rangle
\end{multline}
where we insert into $(\psi^{-},\phi^{+})$ a complete system of 
basis vectors 
\footnote{
We ignore the possible existence of bound states of $H$ of which there
are usually none; certainly not for the $\pi^{+}\pi^{-}$ system of 
$\pi^{+}\pi^{-}\rightarrow \rho^{0}\rightarrow \pi^{+}\pi^{-}$.}
and use 
\begin{equation}
\label{soperator}
\begin{split}
\langle^-\a\,|\,\apr^{+}\rangle
& = (\,\Omega^{-}|\,\a\,\rangle, \Omega^{+}|\,\apr\,\rangle\,)\\
& = \langle\,\a\,|\,\Omega^{-\dagger}
\Omega^{+}\,|\,\apr\,\rangle\\
& = \langle\,\a\,|\,S\,|\,\apr\,\rangle
\end{split}
\end{equation}
Using the invariance of the $S$ operator with respect to space 
time translations
\begin{equation}
\label{translationinvariance}
\left[P_{\mu},S\right]=0
\end{equation}
we obtain
\begin{eqnarray}
\nonumber
\lefteqn{\langle\,\a\eta\,|\,S\,|\,\apr\eta'\,\rangle}\\
\label{reduceds}
& &\qquad\qquad =\delta(\bs{p}-\bs{p'})\delta(p_{0}-p_{0}')
\langle\!\langle\,\a\eta\,|\,\tilde S\,|\,\apr\eta'\,\rangle\!\rangle
\end{eqnarray}
where $\langle\!\langle\quad|\tilde S|\quad\rangle\!\rangle$ is a 
reduced $S$--matrix element. This we can also write as
\begin{multline}
\label{reducedsconvention}
\langle\,\a\eta\,|\,S\,|\,\apr\eta'\,\rangle
=2\hat{E}(\hat{p})\delta(\bs{\hat{p}}-\bs{\hat{p}'})\delta(\sm-\sm')\\
\langle\!\langle\,\a\eta\,|\,S\,|\,\apr\eta'
\,\rangle\!\rangle
\end{multline}
where $\<\!\<\quad|\,S\,|\quad\>\!\>$ is another reduced 
matrix element defined by~\eqref{reducedsconvention}.
In \eqref{reduceds} and
\eqref{reducedsconvention} we include explicitly the degeneracy
quantum number 
$\eta$ for purposes of clarity and completion, 
but we will omit it below for the sake
of notational convenience. 
The form \eqref{reducedsconvention} follows from \eqref{reduceds}
by the defining identities
$\bs{\p}=\frac{\bs{p}}{\sqrt{\sm}}$, 
$\hat{p}^{0}=\frac{p^{0}}{\sqrt{\sm}}$.
From the invariance of the $S$-operator with respect to 
Lorentz transformations,
in particular from $U^{\dagger}(L^{-1}(\hat{p}))S
U(L^{-1}(\hat{p}))=S$~\footnote{$L(\p)$ is the rotation free boost.}
it 
follows that the reduced matrix element
is independent of $\hat{p}$ (i.e., it is the same for 
all $\hat{p}$ as in the center of mass frame
$\bs{\p}=\bs{0}$). Invariance with respect to rotations in the 
center of mass frame shows then by analogous arguments for the discrete 
quantum numbers
$j_{3}$ and $j$ that the reduced matrix element is proportional to
$\delta_{j_{3}j_{3}'}\delta_{jj'}$ and independent of $j_{3}$.
Since Poincar\'e transformations do not change the 
Poincar\'e invariants $\sm$ 
and $j$, the reduced matrix element can still depend upon $\sm$
and $j$. Thus we have
\begin{multline}
\label{thereduced}
\langle\,\a\eta\,|\,S\,|\,\apr\eta'\,\rangle
=2\hat{E}(\hat{p})\delta(\bs{\hat{p}}-\bs{\hat{p}'})\delta(\sm-\sm')
\delta_{j_{3}j_{3}'}\delta_{jj'}\\
\langle\,\eta\,\|\,
S_{j}(\sm)\,\|\,\eta'\,\rangle
\end{multline}
If there are no degeneracy quantum numbers, i.e., if we ignore the 
particle species label and channel numbers and restrict ourselves to the
case without spins (like for the $\pi^{+}\pi^{-}$ system), then the
reduced matrix element can be written as 
\begin{equation}
\label{thereducedelement}
\langle\,\eta\,\|\,S_{j}(\sm)\,\|\,\eta'\,\rangle=S_{j}(\sm)
\end{equation}
where $j$ is the orbital angular momentum in the center of mass. We insert 
\eqref{thereduced} and \eqref{thereducedelement} into
\eqref{smatrixelement} to obtain for the $S$-matrix element
\begin{multline}
\label{amplitude}
(\psi^{-},\phi^{+})=\sum_{j}\int_{(m_{1}+m_{2})^{2}}^{\infty}d\sm
\sum_{j_{3}}\int\frac{d^{3}\hat{p}}{2\hat{E}(\hat{p})}
\langle\,\psi^{-}\,|\,U(L(\hat{p}))\,|\,
\boldsymbol{0}j_{3}j\sm^{-}\rangle
\\S_{j}(\sm)
\langle^{+}\boldsymbol{0}j_{3}j\sm\,|\,U^{\dagger}(L(\hat{p}))
\,|\,\phi^{+}\rangle
\end{multline}
Since resonances come in one partial wave with definite
value of angular momentum, we consider only the one
term in the sum over $j$ with $j=j_{R} (=1^{-}\text{ for }
\pi^{+}\pi^{-}\rightarrow \rho^{0} \rightarrow \pi^{+}\pi^{-})$, 
i.e., we restrict ourselves to the subspace
with $j=j_{R}\,(s=0,\,l=j,\, n=n_{\rho},\,n_{\pi\pi})$. 
This means that we consider only the term with $j=j_{R}$ in the sum
on the right hand side of \eqref{amplitude} and call $S_{j_{R}}(\sm)=S(\sm)$.
To simplify the equations,
we also consider \eqref{amplitude} first for  fixed values of 
$j_{3}$ and $\p$, say $\bs{\p}=\bs{0}$ and $j_{3}=0$. We define
\begin{equation}
\label{ketdefinition}
\begin{split}
\<\psi^{-}(\bs{\p})\,|\,j_{3},\sm^{-}\>&\equiv
\<\psi^{-}\,|\,U(L(\p))\,|\,\bs{0}\,j_{3},\,j=j_{R}\sm^{-}\>\\
&=\overline{\<^{-}\sm, j_{3}\,|\,\psi^{-}(\bs{\p})\>}
\in \tilde{\cal S}\cap {\cal H}_{-}^{2}|_{{\mathbb R}_{\sm_0}}\\
\<^{+}\bs{0}\,j_{3},\,j_{R}\sm\,|\,U^{\dagger}(L(p))\,|
\phi^{+}\>&\equiv\<^{+}j_{3},\sm\,|\,\phi^{+}(\bs{\p})\>
\in \tilde{\cal S}\cap{\cal H}^{2}_{-}|_{{\mathbb R}_{\sm_0}}
\end{split}
\end{equation}
and write~\eqref{amplitude}
\begin{equation}
\label{newamplitude}
(\psi^{-},\phi^{+})=\sum_{j_{3}}\int \frac{d^{3}\p}{2\hat{E}}
(\tilde\psi^{-}_{j_{3}}(\bs{\p}),\tilde\phi^{+}_{j_{3}}(\bs{\p}))
\end{equation}
where
\begin{equation}
\begin{split}
\label{newamplitude2}
\left(\tilde\psi^{-}_{j_{3}}(\bs{\p}),\tilde\phi^{+}_{j_{3}}(\bs{\p})\right)
\equiv\int d\sm \<\psi^{-}\,|\,U(L(\p))\,|\,
\bs{0}j_{3},\, j_{R}\sm^{-}\>S(\sm)\\
\<^{+}\bs{0}j_{3}, \, 
j_{R}\sm\,|\,U^{\dagger}(L(\p))\,|\, \phi^{+}\>\\
=\int d\sm \<\psi^{-}(\bs{\p})\,|\,j_{3}\sm^{-}\>S(\sm)
\<^{+}j_{3}\sm\,|\,\phi^{+}(\bs{\p})\>\, .
\end{split}
\end{equation}
For the fixed values $\bs{\p}=\bs{0}$, $j_{3}=0$ we suppress the labels
$\bs{\p}$ and $j_{3}$ :
\begin{equation}
\label{newamplitude3}
\left(\tilde\psi^{-},\tilde\phi^{+}\right)\equiv
\left(\tilde\psi^{-}_{j_{3}=0}(\bs{\p}=\bs{0}), 
\tilde\phi^{+}_{j_{3}=0}(\bs{\p}=\bs{0})\right)
\end{equation}
and write 
\begin{equation}
\label{newamplitude4}
\left(\tilde\psi^{-}, \tilde\phi^{+}\right)=
\int_{(m_{1}+m_{2})^{2}}^{\infty}d\sm
\<\psi^{-}\,|\,\sm^{-}\>S(\sm)\<^{+}\sm\,|\,\phi^{+}\>\,.
\end{equation}
The quantities $(\tilde\psi^{-},\tilde\phi^{+})$ are not matrix
elements but matrix element densities in the sense that they
have to be calculated for all $4$-velocities $\p$ and all $j_{3}$
and then integrated and summed using \eqref{newamplitude}. Considered
as functions of the invariant mass-squared 
$\sm$, we make the assumption \eqref{t},
i.e., we assume that the
wave functions fulfill 
\begin{equation}
\label{details}
\begin{split}
\!\!\!\!\!
\<^{+}\sm\,|\,\phi^{+}\>
&\in\tilde{\cal S}\cap \H^{2}_{-}|_{{\mathbb R}_{\sm_0}}\\
\!\!\!\!\!
\<^{-}\sm\,|\,\psi^{-}\>&=\overline{\<\psi^{-}\,|\,\sm^{-}\>}
\in\tilde{\cal S}\cap \H^{2}_{+}|_{{\mathbb R}_{\sm_0}}\, ,  
\text{ i.e., }\,\,\<\psi^{-}\,|\,\sm^{-}\>
\in\tilde{\cal S}\cap \H_{-}^{2}|_{{\mathbb R}_{\sm_0}}\,. 
\end{split}
\end{equation}
Then \eqref{newamplitude4} is very much the same as the corresponding
expression for the $S$-matrix element in the non-relativistic
case (e.g., equation (3.9) in \cite{antoniou} or equation (5.4) in \cite{aB97})
except that in place of the non-relativistic energy $E$ with 
lower bound $0$ we have here the center of mass energy squared
$\sm=E^{2}_{cm}$, with lower bound $(m_{1}+m_{2})^{2}$. The 
reason for this choice is that $\sm$ is the variable of the 
relativistic partial $S$-matrix $S_{j}(\sm)$ which is continued
to complex values, and the Riemann energy surface of $S_{j}(\sm)$
has features similar to the Riemann surface of non-relativistic
partial $S$-matrix $S_{l}(E)$. In particular, in the relativistic
$S$-matrix theory $S_{j}(\sm)$ is analytic on the first
``physical'' sheet except for cuts along the real line (and
bound state poles if any exist), and resonances are defined by (pairs of)
poles at the complex values $\sm=\sm_{R}=\left(M_{R}\pm \frac{i}{2}
\Gamma_{R}\right)^{2}$ in the second sheet or another 
``unphysical'' sheet.

\section{Analytic Properties of $\bs{S(\sm)}$}\label{a}

\begin{figure}[p]
\scalebox{.9}{\includegraphics{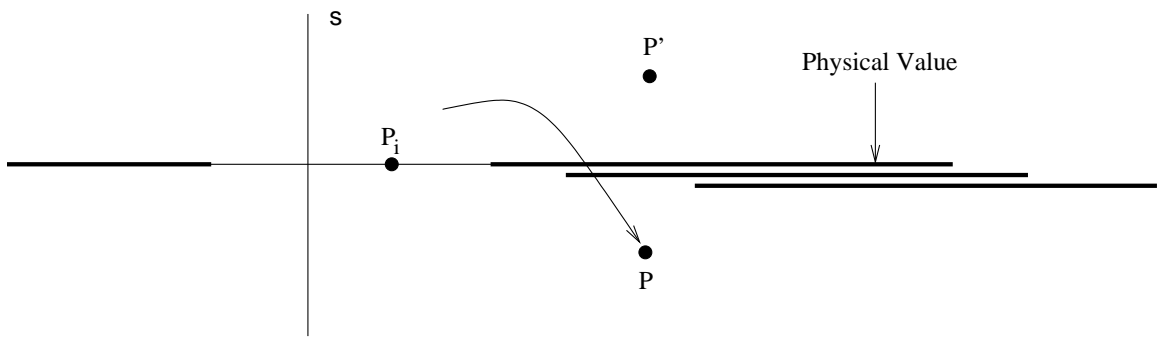}}
\caption{Analytic properties of $S(\sm)$.}\label{d2}
\end{figure}

We start with the hypothesis that unstable particles and 
resonances are associated with poles of the relativistic partial
$S$--matrix $\langle \eta\,\|\, S_{j}(\sm)\,\|\,\eta'\rangle\equiv S(\sm)$
on the ``unphysical'' Riemann sheet. The partial $S$-matrix $S(\sm)$ for
physical values of the center of mass energy $\sm\geq (m_{1}+m_{2})^{2}$ is the
boundary value to the real axis $\sm+i\epsilon\,,\, \epsilon
\rightarrow 0^{+}$ of a function in the complex $\sm$-plane which is analytic 
except for cuts along the real axis and possibly bound state (stable particle)
poles $P_{i}$ at $\sm=m_{P_{i}}^{2}<(m_{1}+m_{2})^{2}$. 
The right hand cut (a consequence of unitarity) starts at the elastic
scattering threshold $\sm_{min}=(m_{1}+m_{2})^{2}$. 
This cut is two-sheeted \cite{zimmermann}, i.e., of 
the square-root type. The elastic
scattering matrix element, $\eta'=\eta$, has further branch points
at each energy $\sm_{th}$ corresponding to a threshold for a newly
allowed 
physical process (e.g., possibly for $\pi^{+}\pi^{-}\rightarrow
\rho \rightarrow \pi^{+}\pi^{-}\pi^{+}\pi^{-}$ at 
$\sm_{th}=(2m_{1}+2m_{2})^{2}$). 
To make $S(\sm)$ single-valued on a Riemann surface,
cuts start at these branch-points and are drawn along the real axis, 
cf.\ Figure~\ref{d2}. These branch-points are called normal thresholds 
and start at energies at which
production of other particles (inelastic processes) is possible.
If one does not cross a cut, one stays on the ``physical sheet''.
The first normal threshold $\sm=(m_{1}+m_{2})^{2}$ is the least $\sm$ at which
a two particle state can exist. To reach other sheets of the Riemann
surface of $S(\sm)$ one burrows through one or several branch cuts.
These sheets are called the ``unphysical'' sheets. The second sheet
is reached from the physical value $\sm+i\epsilon\,,\, \sm>(m_{1}+m_{2})^{2}$
by burrowing down through the normal threshold between $\sm_{th1}
=(m_{1}+m_{2})^{2}$ and $\sm_{th2}$. The resonance poles $P$ 
of the relativistic
elastic scattering $S$-matrix element is located
on the second sheet at $\sm_{R}=(M_{R}-\frac{i}{2}\Gamma_{R})^{2}$.
This hypothesis is suggested by the idea that an unstable particle may
be connected to a stable particle by letting the parameters that measure
the strength of the force (between the decay products) vary continuously
until that force becomes so strong that the unstable particle becomes a 
stable particle, since a stable particle corresponds to a pole on the
physical sheet at the real value $\sm=m_{P_{i}}^{2}$ below the threshold
$(m_{1}+m_{2})^{2}$; in the transition to instability the pole must pass
round the $(m_{1}+m_{2})^{2}$ branch point and through the cut.
Hermitian analyticity (symmetry relation of the $S$--matrix 
$S(\sm-i\epsilon)=S^{*}(\sm+i\epsilon)$) implies that when the pole position
is complex there must be a pole $P'$ at the complex conjugate position
$\sm_{R}^{*}=(M_{R}+\frac{i}{2}\Gamma_{R})^{2}$ on the unphysical sheet
reached by burrowing through the cut from the lower half plane
of the physical sheet of Figure~\ref{d2}.
Thus a scattering resonance is defined by a pair of poles on the second 
sheet of the analytically continued $S$-matrix located at positions
that are complex conjugates of each other.
The pole $P'$ corresponds to the time-reversed situation which we
do not want to discuss here.
(There may exist other resonance poles located at the same or
other physical sheets,
but we will mainly be concerned here with one pair of poles in the
second sheet). 
There may also be higher order poles~\cite{maxson} or 
branch-points on the unphysical sheet 
but we do not venture here into these complications; 
for the sake of simplicity, we shall restrict ourselves to the elastic 
process $\pi^{+}\pi^{-}\rightarrow \pi^{+}\pi^{-}$. 
and assume further that there are no 
other resonance poles in the same channel (ignoring a possible
$\rho(1450)$).

\section{Analytic Extension to the Resonance Pole}\label{continuation}

We want to analytically continue the matrix element density
$(\tilde{\psi}_{j_3}^{-}(\bs{\p}),\tilde{\phi}_{j_{3}}^{+}(\bs{\p}))$
\eqref{newamplitude2} so as to encompass the resonance
pole contribution to the scattering amplitude. As
a specific example, we consider resonance formation in 
elastic scattering, e.g.,  
\begin{equation}
\label{stard}
\pi^{+}\pi^{-}\rightarrow\rho^{0}\rightarrow\pi^+\pi^-\,.
\end{equation}
We will split (see~\eqref{analyticcontinuation} below)
$(\tilde{\psi}_{j_3}^{-}(\bs{\p}),\tilde{\phi}_{j_{3}}^{+}(\bs{\p}))$
into a resonance term and a background term. The resonance
term yields the state vector description of the unstable particle,
which in this case is the $\rho^0$-meson. 
The procedure depicted below can be generalized to processes
other than~\eqref{stard}, with the background term being
determined by the system of branch cuts specific to the particular
process considered.

\begin{figure}[p]
\scalebox{.9}{\includegraphics{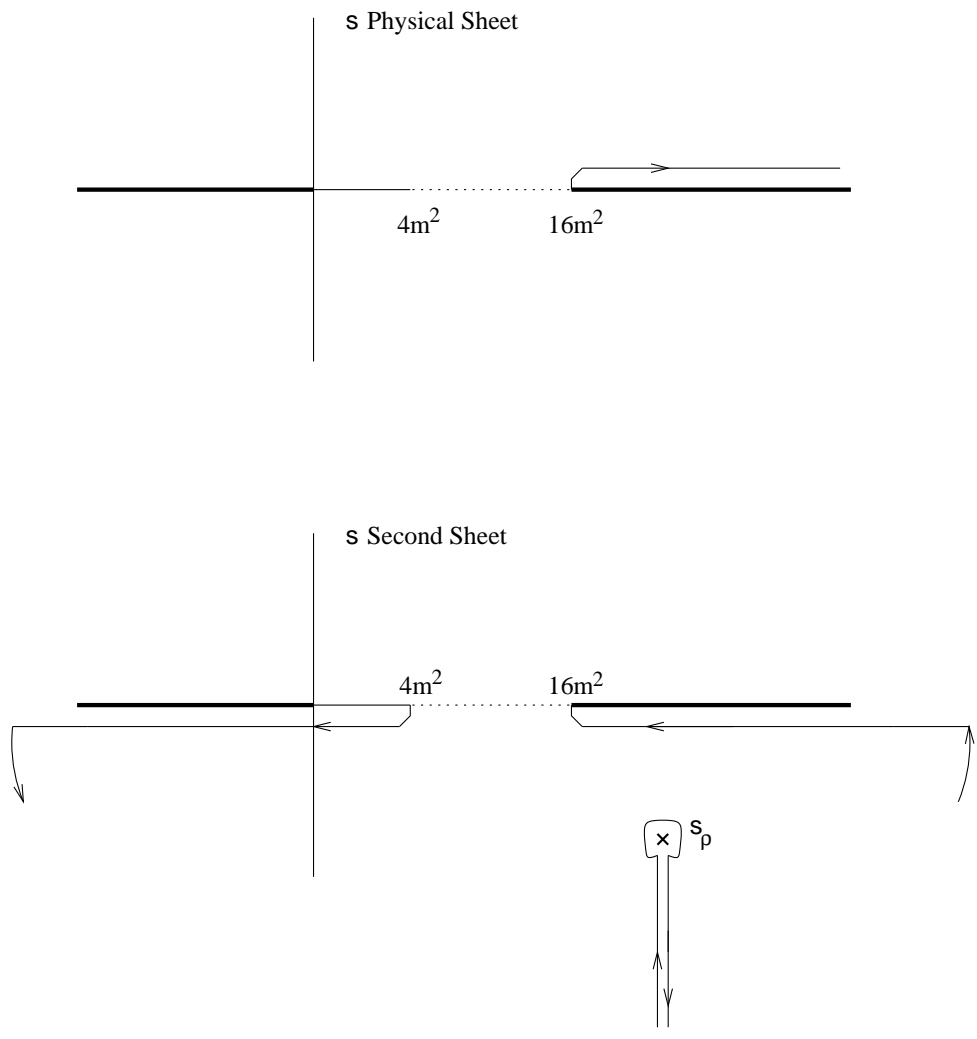}}
\caption{Contour for the analytic continuation of~\eqref{newamplitude2}.}
\label{d1}
\end{figure}

For~\eqref{stard}, 
$\phi^{+}$ is determined from $\phi^{in}$ \eqref{timeevolution}, 
the incoming two pion state  prepared by the
preparation apparatus, and $\psi^{-}$ 
is determined from $\psi^{out}$ \eqref{psiminus},
the registered two pion
state. Since the $\rho$ meson has spin $1$, in \eqref{newamplitude2}
we have $j_{R}=j_{\rho}=1$.
The path of integration in \eqref{newamplitude2} 
extends along the positive real axis,
just above the normal threshold cuts mentioned in 
Section~\ref{a}
from $4m^{2}$ to $\infty$, where $m$ is the mass of $\pi^{+}$. 
As explained in Section~\ref{a}, if we assume 
that the least energetic channel following $\pi^{+}\pi^{-}
\rightarrow \pi^{+}\pi^{-}$ is $\pi^{+}\pi^{-}\rightarrow 
\pi^{+}\pi^{-}\pi^{+}\pi^{-}$, then $S(\sm)$ has a square root branch cut
between $4m^{2}$ and $16m^{2}$. Thus the physical sheet is connected
to the second Riemann sheet between $4m^{2}$ and $16m^{2}$ and $S(\sm)$
can be analytically continued everywhere in the second Riemann sheet
\cite{zimmermann}, except for the resonance pole
of the $\rho^0$ meson. So, to 
analytically extend \eqref{newamplitude2} we deform the contour
of integration between $4m^{2}$ and $16m^{2}$ in the first sheet into
an infinite semi-circle in the lower half plane of the second sheet
(cf.\ Figure~\ref{d1}), taking into account the $\rho$-pole of $S(\sm)$. 
This is the only singularity
of the integrand in \eqref{newamplitude2} since the wave functions
$\<\psi^{-}(\bs{\p})\,|\,j_{3}\sm^{-}\>$ and 
$\<^{+}j_{3}\sm\,|\,\phi^{+}(\bs{\p})\>$,
according to~\eqref{t}, are Hardy class
from below, hence analytic for $\rm{Im}\,\sm<0$. 
Explicitly, the contour deformation of~\eqref{newamplitude2}
yields, according to Figure~\ref{d1},
\begin{eqnarray}
(\tilde\psi_{j_{3}}^{-}(\bs{\p}),\tilde\phi_{j_{3}}^{+}(\bs{\p}))
&=&\int_{4m^{2}}^{\infty}d\sm\<\psi^{-}(\bs{\p})|j_{3}\sm^{-}\>
S(\sm)\<^{+}j_{3}\sm|\phi^{+}(\bs{\p})\>\nonumber\\
&=&\left[\int_{4m^{2}}^{-\infty}d\sm\<\psi^{-}(\bs{\p})|j_{3}\sm^{-}\>
S^{II}(\sm-i\epsilon)
\<^{+}j_{3}\sm|\phi^{+}(\bs{\p})\>\right.\nonumber\\
&&+\int_{{\cal C}_{\infty}} d\sm\<\psi^{-}(\bs{\p})|j_{3}\sm^{-}\>
S^{II}(\sm)\<^{+}j_{3}\sm|\phi^{+}(\bs{\p})\>\nonumber\\
&&+\oint d\sm\<\psi^{-}(\bs{\p})|j_{3}\sm^{-}\>S^{II}(\sm)
\<^{+}j_{3}\sm|\phi^{+}(\bs{\p})\>\nonumber\\
&&\left.+\int_{\infty}^{16m^{2}}d\sm\<\psi^{-}(\bs{\p})|j_{3}\sm^{-}\>
S^{II}(\sm-i\epsilon)\<^{+}j_{3}\sm|\phi^{+}(\bs{\p})\>\right]\nonumber\\
&&\!\!\!+\int_{16m^{2}}^{\infty}d\sm\<\psi^{-}(\bs{\p})|j_{3}\sm^{-}\>
S(\sm+i\epsilon)\<^{+}j_{3}\sm|\phi^{+}(\bs{\p})\>\,.
\label{explicitly}
\end{eqnarray}
In the above expression,
the terms between the brackets $[\quad]$ result from the analytic
continuation of $\int_{4m^{2}}^{16m^{2}}d\sm \cdots$ into the
second Riemann sheet through the square root cut between $4m^{2}$
and $16m^{2}$, $S^{II}$ is the $S$-matrix in the second sheet,
 ${\cal C}_\infty$ refers to the infinite semi-circle
in the lower half-plane of the second sheet, and the integral
$\oint\cdots$ is clockwise around the resonance pole 
$\sm_{R}=(M_R-i\Gamma_R/2)^{2}$ ($\sm_{\rho}=\sm_{R}$) in the second sheet.
For the analytic extension~\eqref{explicitly} to be meaningful, the
integral around ${\cal C}_\infty$ should vanish, and the wave functions,
which are determined for the physical values 
${\mathbb R}_{\sm_0}=[4m^2,\infty)$, should have a unique
extension to the non-physical values of $\sm$, i.e., on 
$(-\infty, 4m^{2})$, which, as can be seen in~\eqref{explicitly}
are always occurring in the second sheet. Since our work follows
closely the derivation of the non-relativistic Gamow vectors, 
to justify~\eqref{explicitly}, we invoke 
below the same arguments used for the non-relativistic
case.

To prove the vanishing of the ${\cal C}_\infty$ integral in~\eqref{explicitly},
we make the same assumption about the growth of the $S$-matrix
as made in~\cite{gadella}, namely the $S$-matrix
on the second sheet is bounded by a polynomial, i.e., for large
${\mathsf s}$, there exists a polynomial $P(\sm)$ such that
$|S_{II}(\sm)|\leq |P(\sm)|$. With this assumption, we
obtain the relation
\begin{eqnarray}
\lefteqn{\int_{{\cal C}_\infty}\left|d\sm\<\psi^{-}(\bs{\p})|j_{3}\sm^{-}\>
S(\sm)\<^{+}j_{3}\sm|\phi^{+}(\bs{\p})\>\right|}\qquad\qquad\nonumber\\
\label{c}
& &\leq\int_{{\cal C}_{\infty}}\left|d\sm\<\psi^{-}(\bs{\p})|j_{3}\sm^{-}\>
P(\sm)\<^{+}j_{3}\sm|\phi^{+}(\bs{\p})\>\right|\,.
\end{eqnarray}
From Proposition~\ref{Property2.2}, it follows that
$P(\sm)\<^{+}j_{3}\sm|\phi^{+}(\bs{\p})\>\in\tilde{\cal S}
\cap{\cal H}_{-}^{2}$.
Hence, a straightforward application of H\"older's inequality
shows that
\begin{equation}
\label{d}
\<\psi^{-}(\bs{\p})|j_{3}\sm^{-}\>
P(\sm)\<^{+}j_{3}\sm|\phi^{+}(\bs{\p})\>\in\H_-^1\,.
\end{equation}
With \eqref{c} and \eqref{d}, the vanishing of the integral
around ${\cal C}_{\infty}$ follows from Corollary~\ref{h:3}.

As mentioned above, the 
wave functions $\<\am|\psi^-\>$ and $\<\ap|\phi^+\>$
should be obtained unambiguously everywhere on the real line of the 
second sheet. The Hardy
class assumptions~\eqref{t} ensure that this
is actually the case. This follows from the
remarkable property that every Hardy class
function is completely determined from its values on a half-axis
of the real line. In other words, there exists a bijective mapping
\begin{equation}
\label{theta}
\theta~:~\Bmp\rightarrow~\Bmp|_{{\mathbb R}_{\sm_0}}\,.
\end{equation}
This result, which follows from a theorem of van-Winter's theorem
(\ref{vanwinter}, Appendix B below)
is as crucial here
as it is for the formulation of the non-relativistic Gamow
vectors. 
The $\theta$
function in~\eqref{theta} allows the 
values of the wave functions on the non-physical region
of the Mandelstam variable $\sm$ on the second sheet to be
uniquely determined from their values on the physical
range $[4m^{2},\infty)$. 

Having elucidated how the Hardy class assumptions~\eqref{t}
provide the mathematical justification of~\eqref{explicitly},
we now consider the integral around the pole separately
$$\oint d\sm\<\psi^{-}(\bs{\p})|j_{3}\sm^{-}\>
S^{II}(\sm)\<^{+}j_{3}\sm|\phi^{+}(\bs{\p})\>\,.$$
For this integral, we expand the partial $S$-matrix $S^{II}(\sm)$ into a 
Laurent series about the resonance pole
\begin{equation}
\label{laurent}
S^{II}(\sm)=\frac{r}{\sm-\sm_{R}}+A(\sm)\, ,
\end{equation}
where $r$ is the residue of $S(\sm)$ at the pole and $A(\sm)$
is an analytic function. The pole term in \eqref{explicitly}
can then be expressed as 
\begin{eqnarray}
\nonumber
\lefteqn{
\oint d\sm\<\psi^{-}(\bs{\p})|j_{3}\sm^{-}\>\<^{+}j_{3}\sm|\phi^{+}(\bs{\p})\>
\frac{r}{\sm-\sm_{R}}}\\
& & =-2\pi i r\<\psi^{-}(\bs{\p})|j_{3}\sm_{R}^{-}\>
\<^{+}j_{3}\sm_{R}|\phi^{+}(\bs{\p})\>\nonumber \\
& &=\int_{-\infty{II}}^{\infty}d\sm\<\psi^{-}(\bs{\p})|j_{3}\sm^{-}\>
\<^{+}j_{3}\sm|\phi^{+}(\bs{\p})\>\frac{r}{\sm-\sm_{R}}\,.
\label{poleterm}
\end{eqnarray}
The first equality in \eqref{poleterm}
the well known theorem of Cauchy and the second equality follows from
a theorem of Titchmarsh (\eqref{a.1}, Appendix B).
With~\eqref{poleterm}, \eqref{explicitly} becomes
\begin{equation}
\label{analyticcontinuation}
\left(\tilde{\psi}^{-}_{j_{3}}(\bs{\p}),
\tilde{\phi}^{+}_{j_{3}}(\bs{\p})\right)
=-2\pi i \,r\,
\<\psi^{-}(\bs{\p})\,|\,j_{3}\sm_{R}^{-}\>
\<^{+}j_{3}\sm_{R}\,|\,
\phi^{+}(\bs{\p})\>+B_{j_{3}}(\bs{\p})
\end{equation}
where $B_{j_{3}}(\bs{\p})$ is given by the other non-zero integrals
in \eqref{explicitly}:
\begin{eqnarray}
B_{j_{3}}(\bs{\p})
&=&\int_{16m^{2}}^{\infty}d\sm\<\psi^{-}(\bs{\p})\,|\,j_{3}\sm^{-}\>
\<^{+}j_{3}\sm\,|\,\phi^{+}(\bs{\p})\>S(\sm+i\epsilon)\nonumber\\
&&-\int_{16m^{2}}^{\infty}d\sm\<\psi^{-}(\bs{\p})\,|\,j_{3}\sm^{-}\>
\<^{+}j_{3}\sm\,|\,\phi^{+}(\bs{\p})\>S^{II}(\sm-i\epsilon)\nonumber\\
\label{backgroundterm}
&&-\int_{-\infty}^{4m^{2}}d\sm\<\psi^{-}(\bs{\p})\,|\,j_{3}\sm^{-}\>
\<^{+}j_{3}\sm\,|\,\phi^{+}(\bs{\p})\>S^{II}(\sm-i\epsilon)\,.
\end{eqnarray}
Since \eqref{analyticcontinuation} is valid for any registered
two pion
state $\psi^{-}_{\pi^{+}\pi^{-}}$, we can omit the 
arbitrary $\psi^{-}_{\pi^+\pi^-}\in\Phi_+$
or $\tilde{\psi}^{-}$ and represent a two pion
in-state density by the functional equation in the space $\Phi_+^\times$:
\begin{equation}
\label{complexbasisexpansion}
\tilde{\phi}^{+}_{j_{3}}(\bs{\p})=-2\pi i\, r
\,|\amd\>\<^{+}j_{3}\sm_{R}\,|\,\phi^{+}
(\bs{\p})\>+|B_{j_{3}}(\bs{\p})\>\, ,
\end{equation}
where
\begin{eqnarray}
|B_{j_{3}}(\bs{\p}) \>&=&\int_{16m^{2}}^{\infty}d\sm
|\amr\>
\<^{+}j_{3}\sm\,|\,\phi^{+}(\bs{\p})\>S(\sm+i\epsilon)\nonumber\\
&&-\int_{16m^{2}}^{\infty}d\sm|\amr\>
\<^{+}j_{3}\sm\,|\,\phi^{+}(\bs{\p})\>S^{II}(\sm-i\epsilon)\nonumber\\
\label{backgroundket}
&&-\int_{-\infty}^{4m^{2}}d\sm |\amr\>
\<^{+}j_{3}\sm\,|\,\phi^{+}(\bs{\p})\>S^{II}(\sm-i\epsilon)\,.
\end{eqnarray}
The expansion \eqref{complexbasisexpansion} is the relativistic version
of the complex basis expansion obtained for the non relativistic
case (cf.\ equation $(5.40)$ of \cite{aB97}).
This is the complex basis expansion if there is one resonance 
in the partial wave. If there are two or $N$ resonances with 
$j=j_R$ at $\sm=\sm_{R_{1}}$, $\sm_{R_{2}}$, $\cdots$ (e.g.,
$\rho(1450)$ and $\rho(770)$ in 
$\pi^+\pi^-\rightarrow  \pi^+\pi^-$), 
then we obtain for each an additional integral 
around the pole $\sm_{R_{i}}$ in \eqref{explicitly}.
For each of these poles separately we follow the above procedure and
obtain, in place of the first term on the right hand side 
of~\eqref{complexbasisexpansion} and~\eqref{4.10.5} below, a sum over
the $\sm_{R_{i}}$ (superposition of interfering resonances).

As prescribed in~\eqref{newamplitude}, the in-state $\phi^+$ of the two pions
(in the subspace $j=j_R$) is obtained from the in-state density
$\tilde{\phi}^+_{j_3}(\bs{\p})$ by integration over the $4$-velocities
and summing over $j_3$. Thus,
\begin{align}
\nonumber
\phi^+&=\sum_{j_3}\int\frac{d^{3}\p}{2\p^0}\tilde{\phi}_{j_3}^{+}(\bs{\p})\\
&=\sum_{j_3}\int \frac{d^{3}\p}{2\p^0}|\amd\>\phi_{j_3}(\bs{\p})+|B\>\,,
\label{4.10.5}
\end{align}
where
\begin{equation}
\label{4.10.6}
\phi_{j_3}(\bs{\p})\equiv -2\pi i r\<^+j_3\sm_{R}|\phi^+(\bs{\p})\>
=-2\pi i r\<^+\bs{\p}j_3[\sm_{R} j_{R}]|\phi^+\>\,,
\end{equation}
and
\begin{equation}
\label{4.10.7}
|B\>=\sum_{j_3}\int\frac{d^3\p}{2\p^0}|B_{j_3}(\bs{\p})\>\,.
\end{equation}
The state vector $|\amd\>$ which appears in~\eqref{complexbasisexpansion}
is the relativistic Gamow vector that we set out to construct.
As apparent from the derivation leading to~\eqref{complexbasisexpansion},
it is obtained from the analytic extension in $\sm$ 
of the Dirac-Lippmann-Schwinger kets $|\amr\>$ to the 
resonance pole in the second sheet of the analytically 
continued $S$-matrix. 
The first term in the right hand side of~\eqref{4.10.5}, which is a 
continuous linear superposition of $|\amd\>$ over the $4$-velocity
$\p$ with fixed values of $\sm_{R}$ and $j_{R}$, represents a velocity
wave packet of the resonance particle. We denote it by
\begin{equation}
\label{4.10.8}
\phi_{j_R \sm_R}^G=\sum_{j_3}\int\frac{d^3\p}{2\p^0}
|\amd\>\phi_{j_3}(\bs{\p})\,.
\end{equation}
We recall that $\phi_{j_3}(\bs{\p})$ in~\eqref{4.10.8}, defined
by~\eqref{4.10.6}, is a Schwartz function with respect to $\bs{\p}$, as
required by~\eqref{t4}.

To obtain the invariant mass-square distribution of $|\amd\>$, we deduce
from the second equality in~\eqref{poleterm} that
\begin{align}
\nonumber
\<\psi^{-}(\bs{\p})|j_{3}\sm_{R}^{-}\>&=
\int_{-\infty_{II}}^{\infty}d\sm
\frac{\<^{+}j_{3}\sm|\phi^{+}(\bs{\p})\>}
{\<^{+}j_{3}\sm_{R}|\phi^{+}(\bs{\p})\>}
\frac{\<\psi^{-}(\bs{\p})|j_{3}\sm^{-}\>}{\sm-\sm_{R}}\\
\label{gv}
&=\int_{-\infty_{II}}^{\infty}d\sm 
\frac{\<\psi^{-}(\bs{\p})|j_{3}\sm^{-}\>}{\sm-\sm_{R}}\, ,
\end{align}
where $-\infty_ {II}$ signifies that the ``unphysical'' values
of $\sm$, $(-\infty, 4m^2)$, occur in the second sheet.
Since~\eqref{gv} is valid for any registered two pion state 
$\psi^{-}_{\pi^{+}\pi^{-}}$, we can represent the state
vector corresponding to the resonance particle by
\begin{equation}
\label{rgv}
|\amd\>
=\frac{i}{2\pi}\int_{-\infty_{II}}^{\infty}
d\sm\frac{|\amr\>}{\sm-\sm_{R}}\, .
\end{equation}
The integral representation in~\eqref{rgv} shows that
the relativistic Gamow vector has a Breit-Wigner distribution
in the $\sm$-variable (invariant mass-squared)
\begin{equation}
\label{bw}
\<\amr|\amd\>\sim\frac{1}{\sm-\sm_R}\,\quad -\infty_{II}<\sm<\infty\,.
\end{equation}
Exactly as in the non-relativistic case
(cf., Proposition~$4$ in~\cite{gadella}),
it can be shown that $|\amd\>$
with the integral representation~\eqref{rgv}
is a continuous antilinear functional on $\Phi_+$, i.e.,
$|\amd\>\in \Phi_{+}^{\times}$. 

The relativistic Gamow vector $|\amd\>$
is a generalized eigenvector of $P^{\mu}$ with 
a complex eigenvalue.
To see this, we use the integral representation~\eqref{poleterm}
with the vector $P_{\mu}\psi^-\in\Phi_+$:
\begin{align}
\nonumber
\<P_\mu\psi^{-}\,|\,\amd\>
&=\frac{i}{2\pi}\int_{-\infty}^{\infty}d\sm\frac{
\<P_\mu\psi^{-}\,|\,\amr\>}
{\sm-\sm_{R}}\\
\nonumber
&=\frac{i}{2\pi}\int_{-\infty}^{\infty}d\sm
\frac{\sqrt{\sm} \p_\mu
\<\psi^{-}\,|\,\amr\>}
{\sm-\sm_{R}}\\
\label{complexeigenvalue}
&=\sqrt{\sm_R}\p_\mu
\<\psi^{-}\,|\,\amd\>\, .
\end{align}
In~\eqref{complexeigenvalue},
we used~\eqref{real-a} to write 
$\<P_\mu\psi^-|\am\>=\sqrt{\sm}\p_\mu\<\psi^-|\am\>$ 
and~\eqref{summary1} to assert that $\sqrt{\sm}\p_\mu\<\psi^-|\am\>$
is a Hardy class function from below, so that
Titchmarsh theorem~\ref{a.1} can be applied to obtain the last equality.
Similarly,
\begin{equation}
\label{complexeigenvalue2}
\begin{split}
\<M^{2}\psi^{-}\,|\,\amd\>
&=\frac{i}{2\pi}\int_{-\infty}^{\infty}d\sm\frac{
\<M^{2}\psi^{-}\,|\,\amr\>}
{\sm-\sm_{R}}\\
&=\frac{i}{2\pi}\int_{-\infty}^{\infty}d\sm
\frac{\sm
\<\psi^{-}\,|\,\amr\>}
{\sm-\sm_{R}}\\
&=\sm_R
\<\psi^{-}\,|\,\amd\>\, .
\end{split}
\end{equation}
Equation~\eqref{complexeigenvalue2}, valid for all $\psi^-\in\Phi_+$,
is the mathematical expression that $|\amd\>$ is a generalized eigenvector
of $M^{2}$ with the complex mass square $\sm_R$ as eigenvalue. This 
is written equivalently as
\begin{equation}
\label{4.10.5.5}
M^{2^{\times}}|\amd\>=\sm_R|\amd\>\,.
\end{equation}
In exactly the same way, the wave-packet~\eqref{4.10.8}
is a generalized eigenvector of the mass square operator
$M^2$ with the eigenvalue $\sm_R$:
\begin{equation}
\label{4.15}
M^{2^\times}\phi^G_{j_R\sm_R}=\sm_R\phi^G_{j_{R}\sm_{R}}\,.
\end{equation}
Hence, $\phi^G_{j_R\sm_R}$ in~\eqref{4.10.8} represents velocity
wave-packets of the unstable particles associated with the pole
$\sm_R$.
\section{Conclusion}
In this paper we have discussed a 
state-vector description of relativistic unstable particles. Following
the norms of particle physics phenomenology, we can anticipate at the
outset that the states of an unstable particle ought to be
characterized by the values of its mass, width and spin, the first two
of which have been combined in several different ways to
a single complex
quantity which is associated with the position of the resonance
pole $\sm_R$ of the relativistic $S$-matrix.
As affirmed by Wigner's classic paper~\cite{wigner},
a stable particle can be given a state vector description where the
state vectors are specified by the generalized eigenvalues of momenta,
real mass and spin. These state vectors in fact furnish an
irreducible unitary representation of the Poincar\'e group. 

In the theory we have proposed in this paper, the state vector
description of unstable particles arises from the relativistic Gamow
vector $|\amd\rangle$,
which are vectors (associated to and) obtained from the pole
term of the $S$-matrix at $\sm=\sm_R$.
It is a simultaneous {\em generalized} eigenvector of the
invariant operators $M^2=P_\mu P^\mu$ and
$\hat{W}=-\hat{w}_\mu\hat{w}^\mu$~\footnote{Here 
$\hat{w}=\frac{1}{2}\epsilon_{\mu\nu\rho\sigma}\hat{P}^{\nu}J^{\rho\sigma}$.}
as well as the four velocity
operators $\hat{P}_\mu$ and the third spin component
$\hat{w}_3$. Evidently, the Gamow vector is a generalized eigenvector
of the momenta as well, an immediate consequence of the defining
identity $P_\mu=M\hat{P}_\mu=\sqrt{{\sm}_R}\hat{P}_\mu$. Thus there
exists  a manifest parallel between Wigner's state vector description of
stable particles and the Gamow vector educed state vector description
of unstable particles. The fundamental difference is the complexness of
the (generalized) eigenvalue of the invariant mass square operator
$M^2$ in the latter case.

The relativistic Gamow vector $|\amd\rangle$ with the complex generalized
eigenvalue $\sqrt{{\sm}_R}$ of the (essentially)
self-adjoint mass operator $M$ was derived here
by analytically extending the Dirac-Lippmann-Schwinger kets
$|\amr\rangle$ in the Mandelstam variable $\sm$ into the lower half
complex plane from its initial range $(m_1+m_2)^2\leq\sm<\infty$,
appertaining to a scattering process of two particles, $(m_1,j_1)$ and
$(m_2,j_2)$. The point $\sm_R$ in the lower half plane at which this
analytic extension yields the Gamow vector $|\amd\rangle$ is 
where there exists a simple pole of the analytic $S$-matrix
designating the unstable particle.

Aside from being an alluring proposition on theoretical grounds, for it
puts our understanding of stable and unstable particles on an equal
footing, this state vector description of unstable particles by Gamow
vectors can be seen to unify various properties heuristically
attributed to unstable particles and resonances. First,
Eqs.~(\ref{gv}) and (\ref{rgv}) 
show that the  Gamow vector $|\amd\rangle$ can be resolved into an
integral representation over a Breit-Wigner distribution
$\frac{1}{\sm-{\sm_R}}$ in the Mandelstam
variable on the second sheet, 
an attribute of its association to the resonance pole at
$\sm_R$. Although not discussed here, it will be shown in the
sequel to this paper that the time
evolution of the Gamow vectors is exactly exponential.
Further,  this exponential 
time evolution will be shown to be given 
by $e^{-\Gamma_R t}$ (with $t$ being the time in the rest frame)
where
$\sm_R=(M_R-i\Gamma_R/2)^2$ is the $S$-matrix pole position.
This implies that the 
width of the Breit-Wigner distribution
of the particle is related to its lifetime $\tau_R$
via the {\em exact} identity
$\Gamma_R=\frac{\hbar}{\tau_R}$; we can then define $M_R$ as the real
mass of the resonance. Thus
the characterization of unstable particles proposed here sheds some
light on the recent debate among particle physics phenomenologists and
theorists on the definition of the mass of the $Z$-boson~\cite{tR98} 
and certain
hadronic resonances~\cite{lC98}.

We shall derive the time evolution of Gamow vectors in the forthcoming
sequel where we shall study the transformation properties of
the Gamow vectors under Lorentz transformations and space-time
translations. This study will show that the linear span of Gamow
vectors $|\amd\rangle,\ \bs{\hat{p}}\in{{\mathbb R}^3}$, furnishes an
irreducible representation of the causal Poincar\'e semigroup, defined
as a semi-direct product of the group of homogeneous Lorentz
transformations with the {\em semigroup} of space-time translations
into the forward light cone. From our requirement that the complexness
of the momenta be solely due to that of mass, i.e., by the required
reality of the velocities $\hat{p}_\mu=\frac{p_\mu}{\sqrt{\sm_R}}$, it
will follow that the (homogeneous) Lorentz subgroup can still be represented
unitarily. However, translations will only form a semigroup, not a
group. This group theoretical study will further support our view that
the characterization of stable and unstable particles 
need not be qualitatively different in that the particles of 
either kind can be specified  according to their
mass and spin by means of irreducible representations of the
Poincar\'e transformations: in the stable case, we have  irreducible
unitary representations of the Poincar\'e group, labeled my real
masses; and in the unstable case, irreducible representations of the
causal Poincar\'e semigroup, labeled by complex masses. The semigroup
of the latter case can be interpreted as representing the asymmetry
of time evolution (say, at rest) of unstable systems and the
irreversible character of decay processes.

\section*{Acknowledgement}
We gratefully acknowledge financial support from the Welch Foundation.
\section*{Appendices}
\appendix
\section{Overview of Rigged Hilbert Space Concepts}
A Rigged Hilbert Space~\cite{gelfand} is the result of the completion
of a scalar product space with respect to three different topologies.
The completion of a vector space with respect to some topology
$\tau$ amounts to including in this space the limit points
of all $\tau$-Cauchy sequences. If one starts with a 
scalar-product  space $\Psi$ and completes it with respect
to the norm induced by the scalar product
$$
\|\phi\|=\sqrt{(\phi,\phi)}\,,
$$
a Hilbert space $\H$ is obtained. On the other hand, if 
$\Psi$ is completed with respect to a topology $\tau_\Phi$
defined by a countable number of norms with some qualifications, 
a countably normed space $\Phi$ is obtained. This countably normed
topology $\tau_\Phi$ is finer than the Hilbert space 
topology $\tau_\H$ so that 
there are more 
$\tau_\Phi$-neighborhoods
than $\tau_\H$-neighborhoods. Hence:
$$
\Psi\subset\Phi\subset\H\,.
$$
A third space of interest is the space of antilinear functionals
on $\Phi$, denoted by $\Phi^\times$. Since $\Phi\subset\H$, it follows
that $\H^\times\subset\Phi^\times$. But, from Hilbert space
theory, $\H=\H^\times$. 
Hence
\begin{equation}
\label{r1}
\Phi\subset\H\subset\Phi^\times\,.
\end{equation}
The triplet~\eqref{r1} is called a Rigged Hilbert Space when 
$\Phi$ is nuclear and dense in $\H$ (with respect to $\tau_\H$).
The fundamental physical axiom of the Rigged Hilbert Space formulation
of quantum physics is that the set of states of the physical system
do not inhabit the entire Hilbert space $\H$ but an appropriately
defined dense subspace $\Phi$ of $\H$. The countably normed topology
of $\Phi$ is constructed so as to yield the algebra of relevant physical
observables continuous as mappings on $\Phi$. It is this feature
of Rigged Hilbert Space theory that is made use of in Section~\ref{setup}
in making the distinction between the set of
prepared states $\Phi_{-}$ and registered observables $\Phi_{+}$
by taking $\Phi_\pm$ as dense subspaces of the same 
Hilbert space $\H$ as in~\eqref{triplets}. This distinction
is what allows semigroup time evolution
 to be incorporated into the quantum mechanical
theory.

The action of an element $F\in\Phi^\times$ on $\phi\in\Phi$,
$F(\phi)$, is denoted--in the Dirac bra-ket notation--by
$$F(\phi)=\<\phi|F\>\,.$$
Since $\H\subset\Phi^\times$, it follows that the Dirac bra-ket
$\<\phi|F\>$ is an extension of the Hilbert space scalar
product in the sense that
$$\<\phi|F\>=(\phi,F)\quad\text{ for }\quad F\in\H\,.$$
The topology on $\Phi^\times$, denoted by $\tau_{\Phi^\times}$,
is the weak$^*$-topology induced by $\Phi$ on $\Phi^\times$.
This means that convergence in $\Phi^\times$ is defined by
\begin{equation}
\label{r2}
F_i\xrightarrow{\tau_{\Phi^\times}}F
\Longleftrightarrow\<\phi|F_i\>\rightarrow\<\phi|F\>\,,\quad
\text{ for all }\phi\in\Phi\,.
\end{equation}

To every $\tau_\Phi$-continuous operator $A$ on
$\Phi$, there corresponds a $\tau_{\Phi^\times}$-continuous operator
$A^\times$ defined on $\Phi^\times$ by 
\begin{equation}
\label{r3}
\<\phi|A^\times F\>\equiv\<A\phi|F\>\,,\quad\text{ for all }
\phi\in\Phi\,,\quad F\in\Phi^\times\,.
\end{equation}
The operator $A^\times$ is called the conjugate 
operator of $A$. It is an extension of the 
Hilbert space adjoint operator $A^\dagger$, since for $F\in\H$ we have
\begin{equation}
\label{r4}
\<\phi|A^\times F\>=(A\phi,F)=(\phi,A^\dagger F)\quad\text{for }
F\in\H\,.
\end{equation}
Hence,
\begin{equation}
\label{r5}
A^\dagger|_{\Phi}\subset A^\dagger\subset A^\times\,.
\end{equation}
It should be stressed that the conjugate operator $A^\times$
can be defined as a $\tau_{\Phi^\times}$-continuous  
operator only when $A$ is a
continuous linear operator on $\Phi$.
In quantum mechanics, it is impossible (empirically) to restrict
oneself to continuous (and therefore bounded) operators $\bar{A}$
in $\H$. However, one can restrict oneself to algebras of observables
$\{A,B,\cdots\}$ described by continuous operators in $\Phi$, if
the topology of $\Phi$ is suitably chosen. Then, $A^\times$, $B^\times$,
$\cdots$ are defined and continuous in $\Phi^\times$.

A generalized eigenvector $|F\>$ of a $\tau_\Phi$-continuous
operator $A$ with a generalized eigenvalue $\omega\in{\mathbb C}$
is defined by the relation
\begin{equation}
\label{r6}
\<A\phi|F\>=\<\phi|A^\times F\>=\omega\<\phi|F\>\,,\quad
\text{ for all }\phi\in\Phi\,.
\end{equation}
Since the vector $\phi$ in~\eqref{r6} is arbitrary, \eqref{r6} can be
formally expressed as
\begin{equation}
\label{r7}
A^\times|F\>=\omega|F\>\,.
\end{equation}
In the Dirac notation the $^\times$ in~\eqref{r7} is suppressed
so that \eqref{r7} reads
\begin{equation}
\label{r8}
A|F\>=\omega|F\>\,.
\end{equation}
If $A$ is a self-adjoint operator, suppressing the $^\times$ 
as in~\eqref{r8} does not lead to confusion since 
$A=A^\dagger\subset\Phi^\times$.
However, if  $A$ is not self-adjoint, a clear distinction between
the operator and its conjugate should be made.
The concept of generalized eigenvectors~\eqref{r7} in  
Rigged Hilbert Space mathematics allows the description of ``eigenstates''
which do not exist in the Hilbert space. For instance, the Dirac scattering 
kets are generalized eigenvectors with eigenvalues belonging to the 
continuous spectrum, and they are not Hilbert space elements. The
Gamow vectors, 
which are used to describe decaying states, are also generalized
eigenvectors which are not in $\H$, 
but, unlike in the case of scattering states, 
their complex eigenvalues do not belong to the Hilbert
space spectrum of the Hamiltonian.
\section{Hardy Class Functions on a Half-plane}\label{h}
\begin{definition}[$\H_\pm^p$ $1\leq p<\infty$]\label{h:1}
A complex function $f(x+iy)$ analytic in the open lower
half  complex plane $(y<0)$ is said to be a Hardy~\cite{hardy}
class function from below of order $p$, $\H_-^p$, if
$f(x+iy)$ is $L^p$-integrable as a function of $x$ for any $y<0$ and
\begin{subequations}
\begin{equation}
\label{h1}
\underset{y<0}{\rm sup}\int_{-\infty}^{\infty}
dx\,\,|f(x+iy)|^{p}<\infty\,.
\end{equation}
Similarly, a complex function $f(x+iy)$ analytic in the open 
upper half complex plane $(y>0)$ is said to be a Hardy
class function from above of order $p$, $\H_+^p$, if
$f(x+iy)$ is $L^p$-integrable as a function of $x$ for all $y>0$, and
\begin{equation}
\label{h2}
\underset{y>0}{\rm sup}\int_{-\infty}^{\infty}
dx\,\,|f(x+iy)|^{p}<\infty\,.
\end{equation}
\end{subequations}
\end{definition}
A property of $\H_\pm^p$ functions is that
their boundary values on the real axis exist almost everywhere and
define an $L^p$-integrable function, i.e.,
if $f\in\H_\pm^p$, then its boundary values 
$f(x)\in L^p({\mathbb{R}})$. Conversely, the values of any
$\H_\pm^p$ function on the upper/lower half-plane are 
determined from its boundary values on the real axis. This result
is provided by a theorem of Titchmarsh:
\begin{theorem}[Titchmarsh theorem]\label{a.1}
If $f\in\H_-^p$,  then 
$$
f(z)=\frac{-1}{2\pi i}\int_{-\infty}^{\infty}
\frac{f(t)}{t-z} dt\,,\quad\text{ for }{\rm Im}\,z<0\,,
$$
and
$$
\int_{-\infty}^{\infty}\frac{f(t)}{t-z}dt=0\,,\quad\text{ for }{\rm Im}\,z>0\,.
$$
Similarly, if $f\in\H_+^p$, then
$$
f(z)=\frac{1}{2\pi i}\int_{-\infty}^{\infty}
\frac{f(t)}{t-z} dt\,,\quad\text{ for }{\rm Im}\,z>0\,,
$$
and
$$
\int_{-\infty}^{\infty}\frac{f(t)}{t-z}dt=0\,,\quad\text{ for }{\rm Im}\,z<0\,.
$$
\end{theorem}
This one-to-one correspondence between the $\H_\pm^p$ functions and
their boundary values on ${\mathbb R}$ allows the identification
of $f(z)$ with $f(x)$ for $f\in\H_\pm^p$.

\begin{figure}[p]
\centerline{\scalebox{.9}{\includegraphics{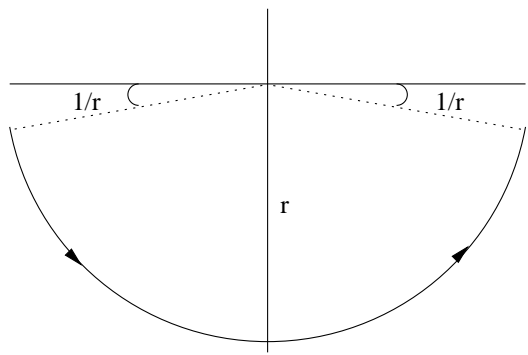}}}
\caption{Arc of radius r, ${\cal C}_{r}$.}\label{arc}
\end{figure}

The following results are related to the decay properties
of the Hardy class functions. They are straightforward
generalizations of the corresponding results of~\cite{gadella}
and are needed for the construction of the relativistic Gamow vectors.
\begin{proposition}\label{h:2}
Let ${\cal C}_{\infty}$ be the infinite semi-circle in the lower
half complex plane. If $f\in\H_-^p$, then
$$
\int_{{\cal C}_\infty}\left|\frac{f(z)}{z} dz\right|=0\,.
$$
\end{proposition}
\begin{proof}
Let ${\cal C}_r$ be the arc with radius $r$ shown in 
Figure~\ref{arc}. Then
$$
\left|\int_{{\cal C}_r}\frac{f(z)}{z} dz\right|\leq 
\int_{{\cal C}_r}|f(r e^{i\theta})|d\theta
=\int_{1/r}^{\pi-1/r}|f(-r e^{i\theta})|d\theta\,.
$$
Since $f\in\H_-^p$, then there exists $C$ such that
$$
|f(-re^{i\theta})|\leq\frac{C}{(r\sin \theta)^{1/p}}\,,
\quad\text{(cf.~\cite{koosis} page 149)}\,.
$$
Thus
\begin{equation}
\label{h3}
\int_{{\cal C}_r}\left|\frac{f(z)}{z} dz\right|\leq 
\frac{2C}{r^{1/p}}\int_{1/r}^{\pi/2}
\frac{1}{(\sin\theta)^{1/p}}d\theta\,.
\end{equation}
Using 
$$\sin\theta\geq \theta-\theta^{3}/6
\geq\theta(1-\pi^2/24)\,,\text{ for }
1/r\leq\theta\leq\pi/2\,,
$$
we obtain for~\eqref{h3}
\begin{eqnarray}
\lefteqn{\int_{{\cal C}_r}\left|\frac{f(z)}{z} dz\right|\leq
\frac{2C}{r^{1/p}(1-\pi^2/24)^{1/p}}
\int_{1/r}^{\pi/2}\frac{d\theta}{\theta^{1/p}}}\nonumber\\
& &\!\!\!\!=
\frac{2C}{(1-\pi^2/24)^{1/p}r^{1/p}}
\begin{cases}
\log\left(\frac{r\pi}{2}\right)
&\,\,\, p=1\\
\frac{1}{1-\frac{1}{p}}
\left[(\frac{\pi}{2})^{1-1/p}-(\frac{1}{r})^{1-1/p}\right]
&\,\,\, 1<p<\infty 
\end{cases}
\end{eqnarray}
Therefore, as $r\rightarrow \infty$, we obtain
$$
\int_{{\cal C}_\infty}
\left|\frac{f(z)}{z} dz\right|=0\,.
$$
\end{proof}
\begin{corollary}\label{h:3}
Let $f\in{\cal S}\cap\H_-^2$,
$g\in{\cal S}\cap\H_-^2$, then
$$
\int_{{\cal C}_\infty}|f(z)g(z) dz|=0\,.
$$
\end{corollary}
\begin{proof}
Since $f\in{\cal S}\cap\H_-^2$, then
$xf(x)\in{\cal S}\cap\H_-^2$~\cite{gadella}. A 
straightforward application of H\"older's inequality
shows that $xf(x)g(x)\in\H_-^1$. Then, from the above lemma
$$
\int_{{\cal C}_\infty}|f(z)g(z)dz|
=\int_{{\cal C}_\infty}\left|\frac{zf(z)g(z)}{z} dz\right|=0\,.
$$
\end{proof}

A remarkable property of Hardy class functions that is
used in~\cite{gadella} is that they are uniquely determined 
from their boundary values on a semi-axis on the real line.
This result is provided by a theorem of van Winter~\cite{winter}. 
Before stating the van Winter's theorem below, we define first
the Mellin transform
\begin{definition}[Mellin transform]
Let $f(x)$ be a function on ${\mathbb R}_{+}$. Its Mellin 
transform is a function defined almost everywhere on ${\mathbb R}$
as
$$
H(s)=\frac{1}{(2\pi)^{1/2}}\int_{0}^{\infty}f(x)x^{is-1/2}dx\,,
$$
provided that the integral exists for almost all $s\in{\mathbb R}$.
\end{definition}
\begin{theorem}[van Winter]\label{vanwinter}
A function $f(x)\in L^{2}({\mathbb R}^+)$ can be extended to
${\mathbb R}^{-}=(-\infty,0]$ to become a function in
$\H_{+}^{2}$ if and only if its Mellin transform satisfies
$$
\int_{-\infty}^{\infty}(1+e^{2\pi s})|H(s)|^{2}ds<\infty\,.
$$
This extension is unique. The values of $f(z)$ for 
$z=\rho e^{i\theta}$ for $0\leq \theta\leq \pi$, $\rho>0$
are given by
$$
f(\rho e^{i\theta})=\frac{1}{(2\pi)^{1/2}}\int_{-\infty}^{\infty}
H(s)(\rho e^{i\theta})^{-is-1/2}ds\,.
$$
In particular for negative values of $x$, $f(x)$ is given by
$$
f(-x)=\frac{1}{(2\pi)^{1/2}}\int_{-\infty}^{\infty}
H(s)(x e^{i\pi})^{-is-1/2}ds\,.
$$
A similar result can be obtained for $\H_{-}^{2}$.
\end{theorem}

\newpage
\end{document}